\documentclass[11pt,a4paper]{article}
\pdfoutput=1 

\usepackage{jheppub}

\DeclareMathOperator\re{\text{Re}}
\DeclareMathOperator\Tr{\text{Tr}}

\newcommand{\be}{\begin{equation}}
\newcommand{\ee}{\end{equation}}
\newcommand{\Ord}{\mathcal{O}}
\newcommand{\ev}[1]{\left\langle #1 \right\rangle}
\newcommand{\SU}{\text{SU}}
\newcommand{\U}{\text{U}}
\newcommand\Nb{N_b}
\newcommand\Nc{N_c}
\newcommand\Nt{N_t}
\newcommand\Tc{T_c}
\newcommand{\bt}{\bar{b}_2}
\newcommand{\btt}{$\bt$}

\title{\boldmath Induced QCD II: Numerical results}

\author[a]{Bastian~B.~Brandt,}
\author[b]{Robert~Lohmayer,}
\author[b]{Tilo~Wettig}

\affiliation[a]{Institute for Theoretical Physics, Goethe University, Max-von-Laue-Strasse 1, 60438 Frankfurt am Main, Germany}
\affiliation[b]{Institute for Theoretical Physics, University of Regensburg, 93040 Regensburg, Germany}
\emailAdd{brandt@th.physik.uni-frankfurt.de}
\emailAdd{robert.lohmayer@ur.de}
\emailAdd{tilo.wettig@ur.de}

\abstract{
  We numerically explore an alternative discretization of continuum
  $\SU(\Nc)$ Yang-Mills theory on a Euclidean spacetime lattice,
  originally introduced by Budzcies and Zirnbauer for gauge group $\U(\Nc)$.
  This discretization can be reformulated such that the self-interactions
  of the gauge field are induced by a path integral over $\Nb$
  auxiliary bosonic fields, which couple linearly to the gauge field.
  In the first paper of the series we have shown that the theory
  reproduces continuum $\SU(\Nc)$ Yang-Mills theory in $d=2$ dimensions
  if $\Nb$ is larger than $\Nc-\frac{3}{4}$ and conjectured, following the argument
  of Budzcies and Zirnbauer, that this remains true for $d>2$. In the
  present paper, we test this conjecture by performing lattice
  simulations of the simplest nontrivial case, i.e., gauge group
  $\SU(2)$ in three dimensions. We show that observables computed
  in the induced theory, such as the static $q\bar q$ potential and
  the deconfinement transition temperature, agree with the same
  observables computed from the ordinary plaquette action up to
  lattice artifacts. We also find that the bound for $\Nb$ can be
  relaxed to $\Nc-\frac{5}{4}$ as conjectured in our earlier paper. Studies
  of how the new discretization can be used to change the order of
  integration in the path integral to arrive at dual formulations of
  QCD are left for future work.
}

\begin{document}

\maketitle

\section{Introduction}

In the strong-coupling limit of lattice gauge theories, gauge fields
do not interact directly with each other, leading to a factorization
of the link integrals in the path integral. This allows both for
analytical investigations and for the construction of new simulation
algorithms
(e.g., \cite{Blairon:1980pk,KlubergStern:1981wz,Kawamoto:1981hw,Rossi:1984cv,Karsch:1988zx}).
Away from the strong coupling limit the self-interactions of the
gauge field need to be taken into account and, with the standard
actions, gauge integrals no longer factorize, spoiling the
applicability of these strong-coupling methods. A particular way to
overcome this problem is to reformulate the gauge action in terms
of auxiliary degrees of freedom so that the gauge fields only couple
to these unphysical degrees of freedom rather than among themselves
(e.g., \cite{Bander:1983mg,Hamber:1983nm,Kazakov:1992ym,Hasenfratz:1992jv}).
In this approach the gauge action is ``induced'' in a well-defined limit
only after the auxiliary degrees of freedom have been integrated out.
Typically this involves taking the limit to an infinite number of fields,
rendering the resulting theories impractical for numerical simulations.

In \cite{Budczies:2003za}, Budczies and Zirnbauer (BZ)
developed a method which induces the pure gauge dynamics already for a
fixed and small number of auxiliary bosonic fields. The key idea is to
give up on the exact reproduction of the Wilson gauge action at finite lattice
spacing in favor of an alternative lattice discretization of Yang-Mills
theory which allows for a formulation in terms of auxiliary bosons.
The vital ingredients for this idea to work are (a) the existence of a
continuum limit and (b) its equivalence with continuum Yang-Mills theory.
For the ``designer action'' (or weight factor) introduced
in~\cite{Budczies:2003za}, BZ could show these properties for gauge group
$\U(\Nc)$ as long as the number of auxiliary bosonic fields, $\Nb$ is
larger or equal to $\Nc$. In QCD we are interested in gauge group $\SU(\Nc)$
and in~\cite{Brandt:2016duy} (the first paper of this series)
we adapted the BZ approach to this case, avoiding a spurious
sign problem in the bosonization of the original formulation by a slight
reformulation of the original weight factor. In
particular, we could show the existence of the continuum limit as
long as $\Nb$ is larger than or equal to $\Nc-1$ (or $\Nc-5/4$ if we allow $\Nb$
to be non-integer) and the equivalence with $\SU(\Nc)$ Yang-Mills theory
in the continuum limit. As in the original BZ paper, the latter could be shown
in $d=2$ dimensions if $\Nb\geq\Nc$ (or $\Nb\geq\Nc-3/4$ if we take $\Nb$ to
be non-integer), but it is a conjecture for $d>2$. A
brief review of the main findings from~\cite{Brandt:2016duy} which are
of importance in this article is included in section~\ref{sec:theory}.

In the present article we will investigate the conjecture numerically
for the simplest nontrivial case, namely $\SU(2)$ gauge theory in
three dimensions. We simulate both the standard theory with the
Wilson plaquette action, which involves a single parameter $\beta$,
and the induced theory with a fixed number of bosonic fields $\Nb$,
which involves a single parameter $\alpha$.  We set the scale for
both theories by computing the Sommer scale $r_0$ \cite{Sommer:1993ce}
from the static quark-antiquark potential.  Matching $r_0$ from both
theories gives us a relation between $\alpha$ and $\beta$.  Using
this relation we can compare other observables, such as quantities
connected to the static $q\bar q$ potential and the
finite-temperature phase transition.  We find that the results agree
very well already away from the continuum limit and that the agreement
improves as the continuum limit is approached.  A preliminary analysis
of data from $\SU(3)$ gauge theory in four dimensions shows that the
modified BZ method also works as expected,
supporting the universality
argument given in~\cite{Budczies:2003za}.
Therefore the modified BZ method, combined with a suitable
strong-coupling approach, can be used to reformulate lattice gauge
theories in a number of different ways. It will be very interesting
to explore such reformulations in the future to see whether they may
have advantages over the traditional formulation (and perhaps even
solve the sign problem afflicting lattice QCD at nonzero density).
We remark that an exact rewriting of the pure gauge action in terms
of auxiliary fields has recently been achieved with
Hubbard-Stratonovich transformations~\cite{Vairinhos:2014uxa}. This
leads to a qualitatively
similar reformulation of the theory, even though the auxiliary fields
are rather different. These two types of reformulations can thus be
seen as complementary approaches with different properties concerning
possible reformulations in terms of dual variables.

This paper is structured as follows.  In section~\ref{sec:theory} we
review the results from the first paper in the
series~\cite{Brandt:2016duy}. The matching between the lattice
couplings in Wilson's pure gauge theory and in the induced gauge
theory is discussed in section~\ref{sec:matching}. In
sections~\ref{sec:qq-num} and~\ref{sec:finiteT} we compare
the results for the static $q\bar{q}$ potential and the deconfinement
phase transition, respectively, before we conclude in
section~\ref{sec:conclusions}. The details concerning the simulation
algorithms and the extraction of the observables as well as the
details of the simulations for comparison between perturbation theory
and numerical results presented in~\cite{Brandt:2016duy} are collected
in the appendixes. First reports of our study have been
published in~\cite{Brandt:2014rca,Brandt:2015nql}.

\section{\boldmath Induced pure gauge theory for gauge group \texorpdfstring{$\SU(\Nc)$}{SU(Nc)}}
\label{sec:theory}

We start by reviewing the results from~\cite{Brandt:2016duy} which
are relevant for this paper. The weight factor of the alternative discretization
of Yang-Mills theory can be written in the form
\be
\label{eq:wfact}
\omega_\text{IPG}[U] = \prod_{p} \Big[\det\big(1-\frac{\alpha}{2}
\big(U_p+U_p^\dagger\big)\big)\Big]^{-\Nb} ,
\ee
where we have adopted the notation used in~\cite{Brandt:2016duy}.
Here, $0\leq\alpha<1$ is the lattice coupling, the index $p$ labels
(unoriented) plaquettes, $U_p$ is the product of link variables
around the plaquette $p$, and $\Nb$ is the number of bosonic fields
in the bosonized version of the weight factor. The weight factor 
\eqref{eq:wfact} is a reformulation of the original weight factor
introduced in~\cite{Budczies:2003za} so that one obtains a real action
after bosonization (cf.~section 2.2 in~\cite{Brandt:2016duy}).
Note that in this formulation of the theory $\Nb$ is just a (positive)
number and thus can take any non-integer value, while the bosonization
is only possible for integer $\Nb$. The weight factor is designed in such
a way that, for a given value of $\Nb$ above the bound in
eq.~\eqref{eq:conti-bound}, the continuum limit is obtained when
$\alpha\to1$.

\begin{figure}
  \centering
  \includegraphics[]{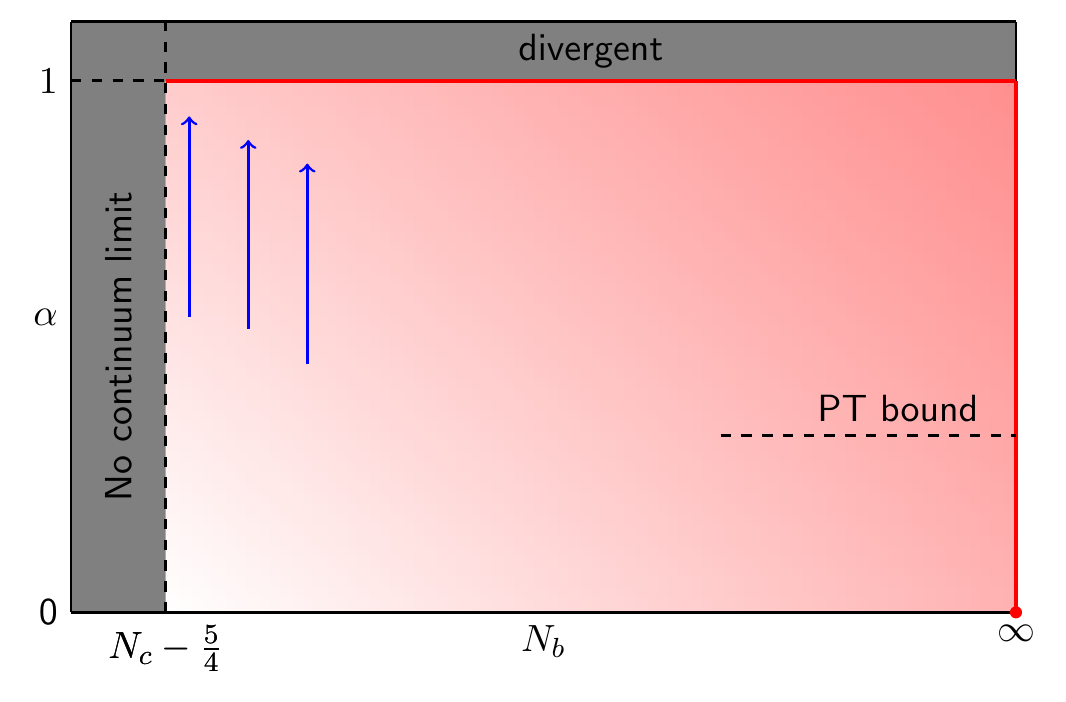}
  \caption{The conjectured phase diagram of $\SU(\Nc)$ induced pure
    gauge theory in the parameter space of the number of bosons $\Nb$
    and the coupling $\alpha$. The darker shading in red indicates
    the approach to the continuum theory, indicated by the red lines.
    The dashed black line in the lower right corner indicates the bound
    $\alpha=1/3$ below which perturbation theory for $\Nb\to\infty$ is valid. The
    blue arrows indicate the interesting region for simulations.}
  \label{fig:ph-diag}
\end{figure}

In~\cite{Brandt:2016duy} we have shown for gauge group $\SU(\Nc)$ that
the theory associated with the weight factor \eqref{eq:wfact},
which we denote as ``induced pure gauge theory'' (IPG),
approaches a continuum limit for $\alpha\to1$ as long as
\be
\label{eq:conti-bound}
\Nb\geq\Nc-\frac{5}{4} \,.
\ee
Furthermore, in two dimensions the theory in the continuum limit is
equivalent to continuum Yang-Mills (YM) theory if
\be
\label{eq:equiv-bound}
\Nb\geq\Nc-\frac{3}{4} \,.
\ee
For $d>2$ the equivalence is a conjecture based on a universality
argument made in~\cite{Budczies:2003za}. As shown in~\cite{Brandt:2016duy},
another way of approaching the continuum limit is to send $\Nb\to\infty$
while keeping $\alpha$ constant. In particular, the lattice theory becomes
equivalent to pure gauge theory with the Wilson plaquette
action~\cite{Wilson:1974sk}
\begin{align}
\label{eq:plaq-act}
  S_W[U] = -\frac{\beta}{\Nc} \sum_p \re\Tr U_p
\end{align}
at lattice coupling $\beta$ when sending $\alpha\to0$ and
$\Nb\to\infty$ while keeping $\beta\sim\Nb\alpha$ fixed. We will refer to the lattice theory with
action \eqref{eq:plaq-act} as ``Wilson pure gauge theory'' (WPG). The 
phase diagram of IPG theory is shown in figure~\ref{fig:ph-diag} in
the parameter space of $\Nb$ and $\alpha$.

In standard discretizations of Yang-Mills theory one can investigate
the nature of the continuum limit by using lattice perturbation theory,
making use of the fact that the bare lattice coupling $g$ goes to zero in
the continuum limit. As noted
in~\cite[sec. 4.1]{Brandt:2016duy}, such an expansion is not possible
in IPG theory around $\alpha=1$ due to the absence of a Gaussian
saddle point. An alternative is to expand the theory around the
saddle point for fixed $\alpha<1/3$ and $\Nb\to\infty$ and to analytically
continue the results to the region where $\alpha\to1$. The associated
bound for perturbation theory is shown as the black dashed line in
figure~\ref{fig:ph-diag}. Using this expansion one finds that a
suitable definition of the coupling in IPG theory in the region
$\alpha\to1$ is given by
\begin{align}
\label{eq:gI}
  \frac1{g_I^2}=d_0(\Nb)\frac{\alpha}{2(1-\alpha)} \,,
\end{align}
where $d_0$ is a perturbative coefficient. The relation between
the couplings $g_W$ and $g_I$ in WPG and IPG, respectively, is then given
by
\begin{align}
\label{eq:coupl-match}
  \frac1{g_W^2}=\frac{1}{g_I^2}\left(1+d_1(\Nb) g_I^2+\ldots\right) 
\end{align}
with another perturbative coefficient $d_1$. The formulas for the
coefficients are given in~\cite[eqs.~(4.12) and (4.13)]{Brandt:2016duy}.

We remark that the weight factor \eqref{eq:wfact} respects
center symmetry, similar to standard Yang-Mills theory and its
discretizations. This is due to the fact that the weight factor is
formulated in terms of the plaquette $U_p$, which itself is a
center-symmetric object. Center symmetry is of fundamental relevance for the
order of the deconfinement transition, which we will study in
section~\ref{sec:finiteT}.

\section{Simulation setup and parameter tuning}
\label{sec:matching}

We would like to test the conjecture that the continuum limit
of IPG reproduces continuum Yang-Mills theory for $d>2$ when the
continuum limit exists, i.e., if eq.~\eqref{eq:conti-bound} is fulfilled.
To this end, we compare results obtained from IPG and WPG while
taking the continuum limit. The first step along the way is to match
the bare parameters such that the simulations in IPG and WPG are done
at similar lattice spacings $a$. In this section we will discuss the
matching for the test case of three-dimensional $\SU(2)$ gauge theory.
This is the computationally cheapest non-trivial case to test the
conjecture. From the theoretical point of view there is nothing
special about this particular case, so that the results obtained
here can be expected to be relevant also for other values of $d>2$ and $\Nc>2$.

\subsection{Simulation setup}

We consider pure $\SU(2)$ gauge theory discretized on a hypercubic
lattice, for which the expectation value of an observable $O$ is
given by
\begin{align}
  \label{eq:obsexp}
  \ev{O}_\text{IPG/WPG} = \frac{1}{Z} \int_{\SU(2)} [dU] \, O[U] \,
  \omega_\text{IPG/WPG}[U] \,.
\end{align}
Here, $Z$ normalizes $\ev1$ to unity, and $\omega[U]$ is the weight factor, which
is given in \eqref{eq:wfact} for IPG and by
\begin{align}
  \omega_\text{WPG}[U] = \exp \big\{ - S_W[U] \big\}
\end{align}
for WPG with the Wilson plaquette action \eqref{eq:plaq-act}.
In IPG a simulation point is characterized by two parameters,
$\Nb$ and $\alpha$, while WPG has only one parameter, $\beta$.

For IPG, we are interested in the continuum limit for a fixed
small value of $\Nb$, i.e., we approach the continuum limit along the
blue arrows in figure~\ref{fig:ph-diag}. In particular, we will perform
the tests for $\Nb=1$ and 2. These two cases are convenient to
test the bounds in eqs.~\eqref{eq:conti-bound}
and~\eqref{eq:equiv-bound} because both cases satisfy the first bound, which
guarantees the existence of a continuum limit, while $\Nb=1$ violates
the second bound, which guarantees equivalence with YM theory in two
dimensions. Following the arguments of section~\ref{sec:theory}, we do not
expect the latter bound to be relevant for $d>2$.

While WPG can be simulated efficiently with the standard combination
of heat-bath~\cite{Kennedy:1985nu} and overrelaxation
updates~\cite{Creutz:1987xi}, such algorithms are not available for
IPG. We thus use a simple Metropolis
algorithm~\cite{Metropolis:1953am} for the simulations, discussed
in detail in appendix~\ref{app:update-algo}.

\subsection{Scale setting and parameter matching}
\label{sec:matching-r0}

We set the scale using the Sommer parameter
$r_0$~\cite{Sommer:1993ce}. This definition of the lattice scale relies
on the physical properties of the static $q\bar{q}$ potential, which
can be computed using Polyakov-loop correlation functions, for instance.
The details of the extraction of the potential $V(R)$ and the Sommer scale
$r_0$ are discussed in appendix~\ref{app:r0-extract}. Since the change from
WPG to IPG amounts to a change of the gauge action only, the operators
relevant for the measurement of the potential, together with their spectral
representation (cf.~appendix~\ref{app:r0-extract}) remain unchanged.

We define equivalent lattice spacings by equivalent values of $r_0/a$.
This means that, for fixed value of $\Nb$, we match the bare parameters
$\beta$ and $\alpha$ so that
$\ev{r_0/a}_\text{WPG}(\beta)=\ev{r_0/a}_\text{IPG}(\alpha)$. When tuned
in such a way, observables are expected to differ by lattice artifacts
only, as long as we are close enough to the continuum. Note that
the resulting functional dependence $\beta(\alpha)$ is not unique.
A similar matching could have been obtained using another observable,
such as the string tension $\sigma$, for instance. The resulting matching
relations will then differ by lattice artifacts.

The strategy for the matching is the following: We start by fitting
the data for $r_0$ obtained from simulations of WPG to the
expression\footnote{This expression is a simple fit ansatz, i.e., it
  contains no assumptions on the relation between $r_0$ and
  $\beta$. We have tested the robustness of the matching obtained from
  this particular choice using several other parameterizations and
  found no significant dependence of the matching coefficients
  in~\eqref{eq:beta_vs_alpha} on the choice of $r_0(\beta)$.}
\begin{align}
  \label{eq:r0-poly}
  r_0(\beta) = \bar{r}_0 + \bar{r}_1 \beta + \bar{r}_2 \beta^2 \,.
\end{align}
Performing a simulation of IPG at a fixed value of $\alpha$ and
computing the corresponding value of $r_0$ then gives us, via
inversion of \eqref{eq:r0-poly}, the value of $\beta$ to which this
particular $\alpha$ should be matched.  This procedure results in a
number of pairs $(\alpha,\beta)$ for a given value of $\Nb$,
which we fit to a suitable
parameterization $\beta(\alpha)$.  The only piece of information we
include in this parameterization is the fact that $\beta\to\infty$
should correspond to $\alpha\to1$.  We write the parameterization
in the form of an asymptotic series,
\begin{align}
  \label{eq:beta-laurent}
  \beta(\alpha) = \sum_{k=-n}^K b_k ( 1 - \alpha )^k\,.
\end{align}
Perturbation theory supports the validity of such an asymptotic
expansion around $\alpha=1$ and suggests $n=1$, see
eqs.~\eqref{eq:gI} and \eqref{eq:coupl-match}.  However, away from
perturbation theory, it is not clear that the asymptotic expansion
still provides a good parameterization of $\beta(\alpha)$.  This
has to be clarified by comparison with the numerical data, and
we shall see below that $n=1$ indeed provides the best description, in
agreement with perturbation theory.

\subsection{Numerical results for the matching}
\label{sec:sim_par}

In order to compare the continuum approach of the two theories we have
performed simulations at four $\beta$-values for WPG for which
high-precision results for the quark-antiquark potential are
available~\cite{Majumdar:2004qx,HariDass:2007tx,Brandt:2009tc,Brandt:2010bw,
  Brandt:2013eua,Brandt:2017yzw,Brandt:2018fft}, i.e., $\beta=5.0$, $7.5$, $10.0$, and
$12.5$. For the simulations of IPG we have estimated the interesting range of
couplings via the expectation value of the plaquette. We have then
chosen three different couplings in this range for $\Nb=1$ and $\Nb=2$
to measure Polyakov-loop correlation functions and the Sommer
parameter. The simulation parameters are given in
table~\ref{tab:simpoints_first}. All statistical error bars quoted in
the following are Jackknife errors obtained with 100 bins. We have
checked explicitly that the error bars and estimates for secondary
quantities do not change significantly if we vary the binsize.

\begin{table}[t]
  \centering
  \small
  \begin{tabular}{cc|crc|ccccccc|l}
    \hline
    \hline
    Theory & \hspace*{-2mm}$\Nb$ & \multicolumn{2}{c}{coupling} & size &
    $R$ & $t_s$ & $n_t$ & $\Delta_\text{sw}$ & meas & $N_\text{sw}$ &
    $\epsilon$ & \multicolumn{1}{c}{$r_0$} \\
    \hline
    \hline
    WPG & & $\beta$ & 5.00 & $24^3$ & 2-10 & 2 & 5k & &
    2000 & & & 3.947(1)(1) \\
    & & & 7.50 & $36^3$ & 2-11 & 2 & 5k & & 2000 & & &
    6.286(7)(1) \\
    & & & 10.00 & $48^3$ & 4-19 & 4 & 5k & & 1800 & & &
    8.603(39)(0) \\
    & & & 12.50 & $64^3$ & 4-19 & 4 & 5k & & 1700 & & &
    10.900(11)(1) \\
    \hline
    IPG & \hspace*{-2mm}1 & $\alpha$ & 0.900 & $24^3$ & 1-10 & 2 &
    100k & 20 & 2000 & 1000 & 0.14 & 3.763(1)(2) \\
    & & & 0.930 & $36^3$ & 1-11 & 4 & 200k & 40 & 2100 & 1500 &
    0.10 & 6.161(2)(1) \\
    & & & 0.945 & $48^3$ & 1-13 & 6 & 500k & 100 & 1700 & 2000 &
    0.08 & 8.363(3)(1) \\
    \hline
    IPG & \hspace*{-2mm}2 & $\alpha$ & 0.650 & $24^3$ & 1-10 & 2 &
    100k & 20 & 2000 & 1000 & 0.14 & 3.329(1)(1) \\
    & & & 0.750 & $36^3$ & 1-11 & 4 & 200k & 40 & 2000 & 1500 &
    0.10 & 5.969(2)(1) \\
    & & & 0.800 & $48^3$ & 1-13 & 6 & 500k & 100 & 2000 & 2000 &
    0.08 & 8.252(2)(1) \\
    \hline
    \hline
  \end{tabular}
  \caption{Simulation parameters and results for the measurements of
    $r_0$ in pure $\SU(2)$ gauge theory with Wilson action
    (WPG) and induced action (IPG) for $\Nb=1$ and 2. Here,
    $R$ gives the range of $q\bar{q}$ separations used in the
    analysis of Polyakov-loop correlation functions, $t_s$ is the temporal
    extent of the L\"uscher-Weisz sublattices, $n_t$ is the number of sublattice
    updates, $\Delta_\text{sw}$ is the number of sweeps separating two
    sublattice measurements, $N_\text{sw}$ is the number of sweeps
    between two measurements, and $\epsilon$ is the size of the ball for the
    link proposal. For more details on the algorithms, e.g.,
    the choice of $\Delta_\text{sw}$ and $N_\text{sw}$, see 
    appendix~\ref{app:sims}.}
  \label{tab:simpoints_first}
\end{table}

The methodology for the extraction of $r_0$ introduced in
appendix~\ref{app:r0-extract} relies on the assumption that IPG is
a confining gauge theory for $\alpha<1$. This is not guaranteed, but
the existence of a minimal coupling after which the theory is confining
in the approach to the continuum limit is a necessary criterion for the
approach to continuum Yang-Mills theory. The simulations have shown that
this is the case for all couplings of table~\ref{tab:simpoints_first} so
that we can extract $r_0$ and use it for scale setting. The results for
$r_0$ are also listed in table~\ref{tab:simpoints_first}. The first
error is statistical, and the second error is the uncertainty associated
with the interpolation.  Note that we have kept the volume at
$L/r_0\gtrsim7$ in order to ensure that finite size effects for $r_0$
are small. That this is indeed the case can be seen from a comparison
of the data presented in table~\ref{tab:simpoints_first} and the
results for $r_0$ given in~\cite{Brandt:2013eua,Brandt:2017yzw}, where
$L/r_0\gtrsim10$. The different results are in very good agreement
within the statistical accuracy.

\begin{table}[t]
  \centering
  \small
  \begin{tabular}{l|lll|c}
    \hline
    \hline
    Fit & \multicolumn{3}{c|}{parameters} & $\chi^2/$dof \\
    \hline
    \hline
    WPG eq.~\eqref{eq:r0-poly} & $\bar{r}_0$ & $\bar{r}_1$ &
    $\bar{r}_2$ & \\
    \hline
    & -0.79(3) & 0.957(9) & -0.0017(6) & 0.01 \\
    \hline
    \hline
    \vspace*{-4mm} & & & & \\
    IPG eq.~\eqref{eq:mfit-1} & $b$ & $b_0$ & $n$ \\
    \hline
    $\Nb=1$ & 0.54(\phantom{2}5) & -1.0(2) & 1.00(1) & $\diagup$ \\
    $\Nb=2$ & 2.51(24) & -2.8(4) & 0.99(5) & $\diagup$ \\
    \hline
    \hline
    IPG eq.~\eqref{eq:mfit-2} & $b_{-1}$ & $b_0$ & $b_1$ & \\
    \hline
    $\Nb=1$ & 0.623(\phantom{1}4) & -1.78(11) & $\:$3.6(7) & $\diagup$ \\
    $\Nb=2$ & 2.453(14) & -2.62(12) & -0.1(3) & $\diagup$ \\
    \hline
    \hline
    IPG eq.~\eqref{eq:mfit-2}; $b_1=0$ & $b_{-1}$ & $b_0$ & & \\
    \hline
    $\Nb=1$ & 0.602(1) & -1.218(\phantom{1}9) & & 109 \\
    $\Nb=2$ & 2.463(4) & -2.694(14) & & 2.2 \\
    \hline
    \hline
  \end{tabular}
  \caption{Results for the parameters of the matching fits (see
    text).  If $\chi^2/$dof is not given, the parameterizations
    contain as many parameters as data points.}
  \label{tab:matching-res}
\end{table}

We start the matching of the two theories by fitting the WPG data
to the form given in eq.~\eqref{eq:r0-poly}. The results
are given in table~\ref{tab:matching-res}.
Note that we have added statistical and systematic uncertainties
in quadrature when fitting the data, which likely overestimates the
true uncertainties and can thus explain the rather small value of
$\chi^2/$dof in the fit.
Using these results as a
definition for the behavior of $r_0$ with $\beta$, we then try to
find a matching between $\beta$ and $\alpha$ which leads to identical
values of $r_0$ in the two theories. The guideline for the
functional form of $\beta(\alpha)$ is
eq.~\eqref{eq:beta-laurent}. First, it is necessary to determine the
leading-order of the divergence of $\beta$ in the limit
$\alpha\to1$. To this end the data for $r_0$ obtained in IPG are
parameterized by eq.~\eqref{eq:r0-poly} and the
parameters from table~\ref{tab:matching-res}, with $\beta$ replaced
by
\begin{align}
  \label{eq:mfit-1}
  \beta(\alpha) = \frac{b}{(1-\alpha)^n} + b_0 \,,
\end{align}
where $b$, $b_0$, and $n$ are free parameters. Note that
this is not a fit, since there are as many free parameters as data
points. The results for the parameters are listed in
table~\ref{tab:matching-res}. The important information is that in
both cases, $\Nb=1$ and $\Nb=2$, we have $n\approx1$ to good accuracy,
implying that we can expect the divergence to be a simple pole. This
suggests that a suitable parameterization of $\beta(\alpha)$ is given
by
\begin{align}
  \label{eq:mfit-2}
  \beta(\alpha) = \frac{b_{-1}}{1-\alpha} + b_0 + b_1(1-\alpha) + \ldots
\end{align}
We test this parameterization by comparison to the data, using either
all three terms (in which case there is no fit) or setting $b_1=0$ (in
which case we have a fit with one degree of freedom). The resulting
parameters are also listed in table~\ref{tab:matching-res}.
We see that the parameters with $b_1\ne0$ and $b_1=0$ are in good
agreement for $\Nb=2$ while there are some deviations for $\Nb=1$. It appears
that this is due to the large correction of the term including $b_1$, which
also leads to a large value of $\chi^2/$dof for the fit where $b_1=0$ with
$\Nb=1$. Even though a $\chi^2/$dof of 2.2 is not satisfactory, the fit for the
$\Nb=2$ case, in contrast, works reasonably well.  Since in both cases the
correction term including $b_1$ is not negligible (for $\Nb=2$ signaled by
$\chi^2/\text{dof}>2$) we take the parameterization with $b_1\ne0$ for the
matching between $\beta$ and $\alpha$,
\begin{subequations}
  \label{eq:beta_vs_alpha}
  \begin{align}
    \beta(\alpha) & =
    \frac{0.623(\phantom{1}4)}{1-\alpha} - 1.78(11) + 3.6(7) (1-\alpha) \quad\text{for
}\Nb=1\,,\\
    \beta(\alpha) & = 
    \frac{2.453(14)}{1-\alpha} - 2.62(12) - 0.1(3) (1-\alpha) \quad\text{for
}\Nb=2\,.
  \end{align}
\end{subequations}
This relation will be updated in the next section, see
eq.~\eqref{eq:beta_vs_alpha_upd} below.  Figure~\ref{fig:r0_matching} shows the
data for $r_0$ vs $\beta$, including also the data from the induced theory
with $\beta$ obtained from \eqref{eq:beta_vs_alpha}.

\begin{figure}[t]
  \centering
  \includegraphics[]{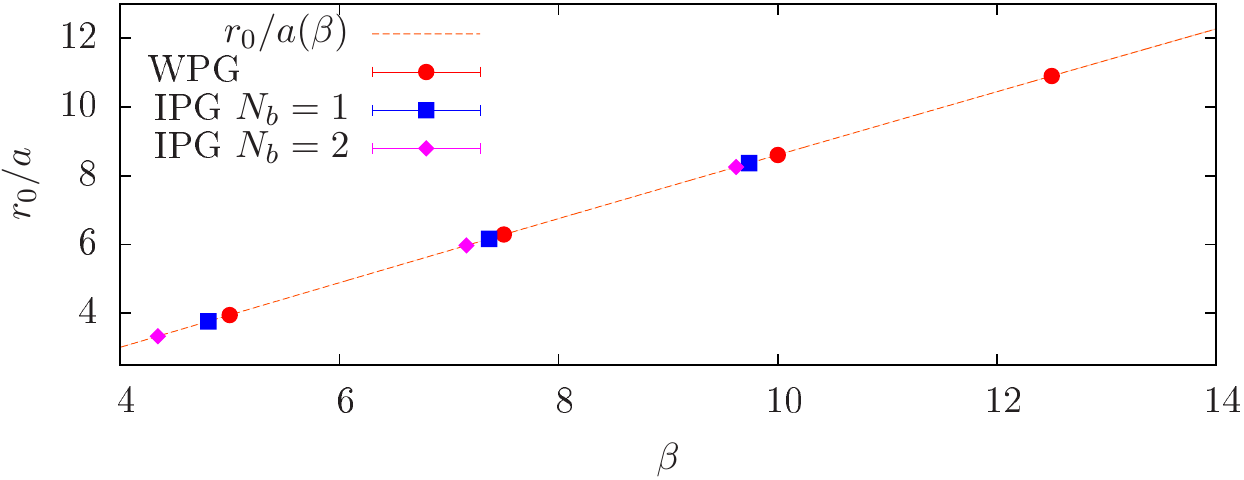}
  \caption{Results for $r_0$ vs $\beta$. The $\beta$-values
    associated with the couplings $\alpha$ for the induced theory have
    been obtained using \eqref{eq:beta_vs_alpha}, which defines the
    matching $\beta(\alpha)$. The red curve is the interpolating
    curve \eqref{eq:r0-poly} with the coefficients from
    table~\ref{tab:matching-res}.}
  \label{fig:r0_matching}
\end{figure}

The results from this type of matching can also be compared to the results from
perturbation theory obtained in~\cite{Brandt:2016duy}. The comparison between the
leading-order coefficient $b_{-1}$ to the perturbative coefficient has already
been done in section 4.3 of~\cite{Brandt:2016duy} and shows an excellent agreement
between numerical data and the perturbative prediction, surprisingly even for
$\Nb=1$. The details of the comparison and the results for larger values of $\Nb>2$
are summarized in appendix~\ref{app:matching-pert}.

\section[The static \texorpdfstring{\boldmath $q\bar q$}{qq}
potential]{\boldmath The static $q\bar q$ potential}
\label{sec:qq-num}

After matching the bare couplings of WPG and IPG we are now
in a position to compare the results for other observables.
In this section we will focus on the static $q\bar{q}$ potential,
which is not only important to show that the theory is confining
but is also related to an effective string theory for the
QCD flux tube (for a recent review see~\cite{Brandt:2016xsp}).
The latter contains non-universal parameters, which can be
used to distinguish between different microscopic theories.

\subsection{Effective string theory and analysis strategy}

A possible way to analyze the static $q\bar{q}$ potential is
a comparison to the predictions from the associated effective
low-energy theory, namely the effective string theory (EST) for
the QCD flux tube, which governs the potential at intermediate and
large $q\bar{q}$ separations $R$. The interested reader is referred
to the reviews~\cite{Aharony:2013ipa,Brandt:2016xsp} for
further reading about the foundations of the EST and a thorough
list of references.

The main result from the EST that we will use in the analysis
is the prediction for the $R$ dependence of the static
$q\bar{q}$ potential for large
$R$~\cite{Arvis:1983fp,Aharony:2010db,Aharony:2011ga,
Dubovsky:2012sh,Dubovsky:2014fma},
\be
\label{eq:est-spec}
V^\text{EST}(R) = \sigma R \, \sqrt{ 1 - \frac{\pi\:(d-2)}{12\sigma R^{2}} }
 - \bt \frac{\pi^3(d-2)}{60 \sqrt{\sigma^3} R^4}
 + \Ord(R^{-\xi}) \,,
\ee
where $\sigma$ is the string tension and $\xi=6$ or
$\xi=7$, depending on whether the next term in the large-$R$ expansion
originates from another boundary term or a bulk term.
The first term on the right-hand side is the light-cone (LC)
spectrum~\cite{Arvis:1983fp}, which is expected to appear due to the
integrability of the leading-order $S$-matrix in the analysis using the
thermodynamic Bethe ansatz~\cite{Dubovsky:2012sh,Dubovsky:2014fma}. The
appearance of the full square-root formula is also in good agreement
with numerical results for the potential (see~\cite{Brandt:2016xsp} for
a compilation of results). The parameter \btt{} in eq.~\eqref{eq:est-spec}
is the leading-order boundary
coefficient~\cite{Aharony:2010db,Aharony:2011ga}, which has been found
to be non-universal~\cite{Brandt:2017yzw,Brandt:2018fft}. In the spectrum, possible
corrections to the standard EST energy levels, such as the rigidity
term, first proposed by Polyakov~\cite{Polyakov:1996nc}, and corrections
due to massive modes~\cite{Dubovsky:2012sh,Dubovsky:2014fma}, have been
left out.

Our basic strategy is to try to reproduce, and to compare to, the
high-precision results from~\cite{Brandt:2017yzw}, performing the same
analysis steps. Since our aim is to perform a like-by-like comparison,
rather than validating the string picture, we focus on the basic EST
analysis, i.e.,
sections~3 and 4 in~\cite{Brandt:2017yzw}.

\subsection{Simulation points and scale setting}
\label{sec:sim-points}

For the comparison we have performed simulations at bare parameters $\alpha$
and volumes that are matched to the parameters in~\cite{Brandt:2017yzw} using
the matching relations \eqref{eq:beta_vs_alpha}. The new set
of parameters is shown in table~\ref{tab:simpoints_second}.

\begin{table}[t]
  \centering
  \small
  \begin{tabular}{cc|cccc|cccccc|l}
    \hline
    \hline
    Lattice & $\Nb$ & $\alpha$ & $\beta^\text{ref}$ & size &
    $R$ & $t_s$ & $n_t$ & $\Delta_\text{sw}$ & \#meas & $N_\text{sw}$ \\
    \hline
    \hline
    I$^1_1$ & 1 & 0.903 & 5.00 & $48^3$ & 1-10 & 2 & 400k & 20 & 2400 & 2000 \\
    I$^1_2$ & 1 & 0.931 & 7.50 & $64^3$ & 1-15 & 4 & 800k & 40 & 2200 & 3000 \\
    I$^1_3$ & 1 & 0.946 & 10.0 & $96^3$ & 1-22 & 6 & 2000k & 100 & 1900 & 6000
\\
    \hline
    I$^2_1$ & 2 & 0.680 & 5.00 & $48^3$ & 1-10 & 2 & 400k & 20 & 2600 & 2000 \\
    I$^2_2$ & 2 & 0.759 & 7.50 & $64^3$ & 1-15 & 4 & 800k & 40 & 2200 & 3000 \\
    I$^2_3$ & 2 & 0.806 & 10.0 & $96^3$ & 1-22 & 6 & 2000k & 100 & 2000 & 6000
\\
    \hline
    \hline
  \end{tabular}
  \caption{Simulation parameters for the high-precision measurements of
    the static $q\bar{q}$ potential in simulations with the induced
    action.  $\beta^\text{ref}$ is the target value of
    $\beta$ in WPG theory.  The meaning of the other parameters is the
    same as in table~\ref{tab:simpoints_first}.  To avoid the propagation of rounding
    errors, the $\alpha$-values were computed using all digits of $b_{-1}$,
    $b_0$ and $b_1$ instead of the rounded values in \eqref{eq:beta_vs_alpha}.}
  \label{tab:simpoints_second}
\end{table}

For the purpose of scale setting and to check the matching of
eq.~\eqref{eq:beta_vs_alpha} we have computed the Sommer
parameter using the same strategy as before. The results are given
in table~\ref{tab:sim_results_qq}. We can use these results to update the
scaling relation~\eqref{eq:beta_vs_alpha}. For the parameterization
with $b_1\neq0$ we obtain
\begin{subequations}
  \label{eq:beta_vs_alpha_upd}
  \begin{align}
    \beta(\alpha) & =
    \frac{0.618(\:\:3)}{1-\alpha} - 1.63(\:\:9) + 2.5(7) (1-\alpha)
\quad\text{for
}\Nb=1\,,\\
    \beta(\alpha) & = 
    \frac{2.439(15)}{1-\alpha} - 2.48(13) - 0.5(3) (1-\alpha) \quad\text{for
}\Nb=2\,.
  \end{align}
\end{subequations}

\begin{table}[t]
  \centering
  \small
  \begin{tabular}{c|l|l|cll}
    \hline
    \hline
    Lattice & \multicolumn{1}{c|}{$r_0$} & 
    \multicolumn{1}{c|}{$\langle U_p \rangle/\Nc$} & $R_\text{min}$ &
    \multicolumn{1}{c}{$\sqrt{\vphantom{\sigma}\smash{\sigma_{(i)}}}\,r_0$} &
    \multicolumn{1}{c}{$\sqrt{\vphantom{\sigma}\smash{\sigma_{(ii)}}}\,r_0$} \\
    \hline
    \hline
    I$^1_1$ & 3.9264(5)(39) &  0.795094(16) & 4 & 1.2292(17) & 1.2309(5)
    \\
    I$^1_2$ & 6.2804(6)(\phantom{3}3) &  0.865488(\phantom{1}8) & 5 & 1.2398(34) & 1.2355(8)
    \\
    I$^1_3$ & 8.5527(6)(\phantom{3}8) &  0.898656(\phantom{1}5) & 8 & 1.2384(26) & 1.2357(6)
    \\
    \hline
    I$^2_1$ & 3.9401(3)(44) &  0.790699(\phantom{1}9) & 4 & 1.2340(28) & 1.2320(6)
    \\
    I$^2_2$ & 6.3121(5)(\phantom{3}3) &  0.863568(\phantom{1}5) & 5 & 1.2342(38) & 1.2340(8)
    \\
    I$^2_3$ & 8.6057(7)(\phantom{3}8) &  0.898044(\phantom{1}4) & 8 & 1.2347(27) & 1.2346(7)
    \\
    \hline
    \hline
  \end{tabular}
  \caption{Results for the Sommer parameter $r_0$, the average plaquette
    $\langle U_p \rangle$, the string tension $\sigma$ as obtained from
    methods (i) and (ii) described in the text, for the simulations listed
    in table~\ref{tab:simpoints_second}.}
  \label{tab:sim_results_qq}
\end{table}

\begin{figure}[t]
  \centering
  \includegraphics[]{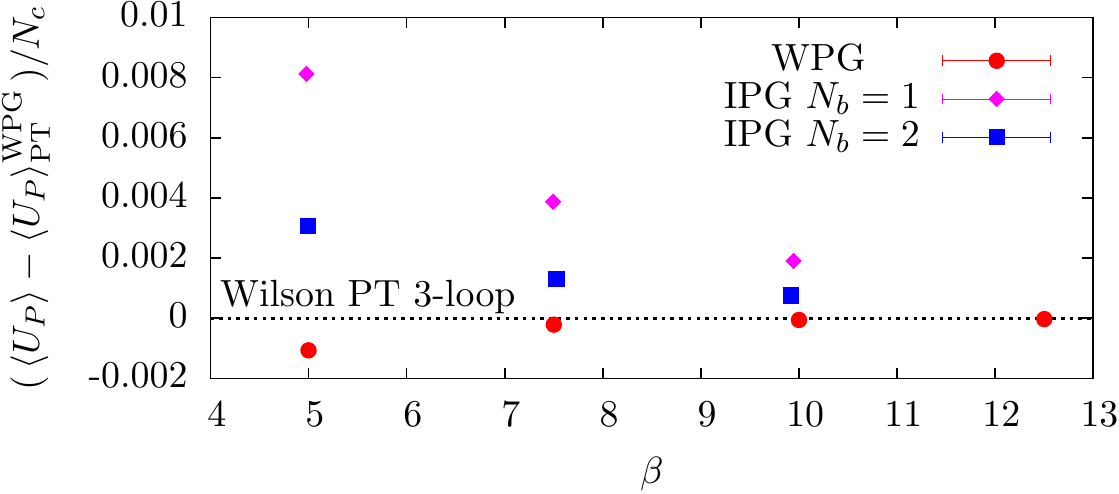}
  \caption{Results for $\langle U_p \rangle/\Nc$ relative to the
    perturbative result for the Wilson action of
    eq.~\eqref{eq:plaquette_pert} vs $\beta$.  The $\beta$-values
    associated with the couplings $\alpha$ for the induced theory have
    been obtained using eq.~\eqref{eq:beta_vs_alpha_upd}.}
\label{fig:plaquette}
\end{figure}

For a first comparison of observables in the approach to the continuum
we can look at the expectation value of the plaquette. Even though
trivial in the continuum limit, its dependence on the lattice spacing,
i.e., the bare coupling, is non-trivial and can be computed in lattice
perturbation theory. The three-loop result for $\SU(2)$ WPG in $d=3$
is given by~\cite{Panagopoulos:2006ky}
\begin{align}
  \label{eq:plaquette_pert}
  \langle U_p \rangle^\text{WPG}_\text{PT}/\Nc
  = 1 - 0.25 g_0^2 - 0.01453916(4) g_0^4 - 0.0053459(2) g_0^6 \,.
\end{align}
The numerical results for the plaquette are listed in
table~\ref{tab:sim_results_qq} and shown in
figure~\ref{fig:plaquette}. To make the small differences visible we do
not display the raw data but rather the difference between the plaquette
expectation values and the perturbative result.

The plot shows that the WPG results are already very close to the
perturbative result, i.e., lattice artifacts with respect to 3-loop
perturbation theory for Wilson's gauge action are small.  From the plot
one might be led to the conclusion that lattice
artifacts for IPG are larger. However, the coefficients of the perturbative
result which we subtracted have been obtained from Wilson's plaquette action and
are expected to be different for IPG in general, and for different values
of $\Nb$ in particular.  Indeed, corrections to eq.~\eqref{eq:plaquette_pert}
in WPG theory start at order $g_0^8$, while for IPG theory corrections start
at lower orders.

Another option to set the scale is via the string tension $\sigma$, which
governs the linear rise with $R$ of the potential for $R\to\infty$. As
in~\cite{Brandt:2017yzw} we extract $\sigma$ with two different methods:
\begin{enumerate}
 \item[(i)] We fit the force to the form 
 \be
  \label{eq:force-sig-fit}
  R^2 F(R) = \sigma R^2 + \gamma \,,
 \ee
 motivated by the expansion of the EST potential to  next-to-leading-order
 in $1/R$.
 \item[(ii)] We fit the potential to
 \be
  \label{eq:LC-pot}
   V(R) = \sigma R \, \sqrt{ 1 -
   \frac{\pi(d-2)}{12\sigma R^{2}} } + V_0 \,,
 \ee
 corresponding to the (full) leading-order EST prediction with an additive
 normalization constant $V_0$.
\end{enumerate}
This particular combination of ans\"atze is especially useful since corrections
with respect to the full EST prediction, eq.~\eqref{eq:est-spec}, appear at
different orders in the $1/R$ expansion. Consequently, we can determine the
region where higher-order terms in the EST are negligible by comparing the
results for $\sigma$ from methods (i) and (ii). The basic strategy is to
investigate the dependence of $\sigma$ on the minimal value of $R$ included in
the fit, $R_\text{min}$.
In particular, in the region where the results from the two methods agree
within errors and show a plateau, higher-order terms are expected to be negligible,
and we can use any of the results for $\sigma$.

\begin{figure}[t]
  \centering
  \includegraphics[]{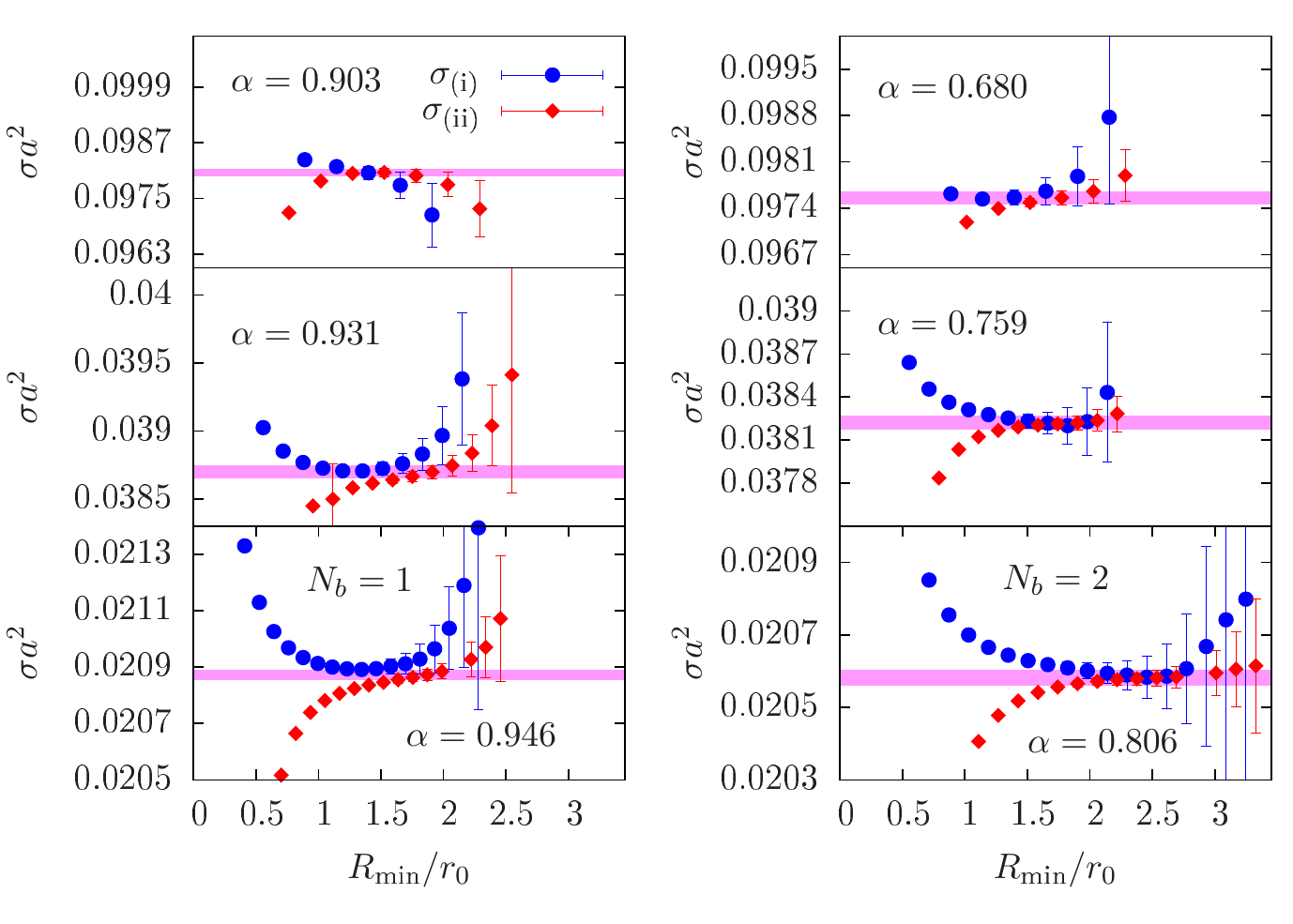}
  \caption{Results for the string tension in IPG for $\Nb=1$ (left) and $\Nb=2$
  (right) extracted from method (i), $\sigma_\text{(i)}$, and (ii),
  $\sigma_\text{(ii)}$, as explained in the text, vs the minimal value of $R$
  included in the fit, $R_\text{min}$ in units of $r_0$. The red bands are the
  values of $\sigma_\text{(ii)}$ which we use for the further analysis.}
\label{fig:sigma_vs_Rmin}
\end{figure}

In figure~\ref{fig:sigma_vs_Rmin} we show the results for the extraction of the string tension in IPG for $\Nb=1$
(left) and $\Nb=2$ (right). We see that for
$\Nb=1$ and $\alpha=0.931$ and 0.946 the results from method (i) already start
to leave the plateau region when the results from method (ii) reach the plateau.
Nevertheless, the plateau values agree within uncertainties, as the results from
method (i) agree within errors at the point where the results from method (ii)
reach the plateau region.
As the final results for the string tension, we will use the result
from method (ii) obtained with the value of $R_\text{min}$ where the results
of the two methods become consistent. In the cases without a common plateau we
use the value of $R_\text{min}$ where the result from method (ii) starts to
agree with the plateau from method (i). The final results are indicated by the
red bands in figure~\ref{fig:sigma_vs_Rmin}.
The results for $\sigma$ are listed in
table~\ref{tab:sim_results_qq}. The quoted uncertainties for $\sigma$ include the
systematic uncertainties for $r_0$, which have been conservatively added in
quadrature to the statistical uncertainties.

\begin{figure}[t]
 \centering
\includegraphics[width=14cm]{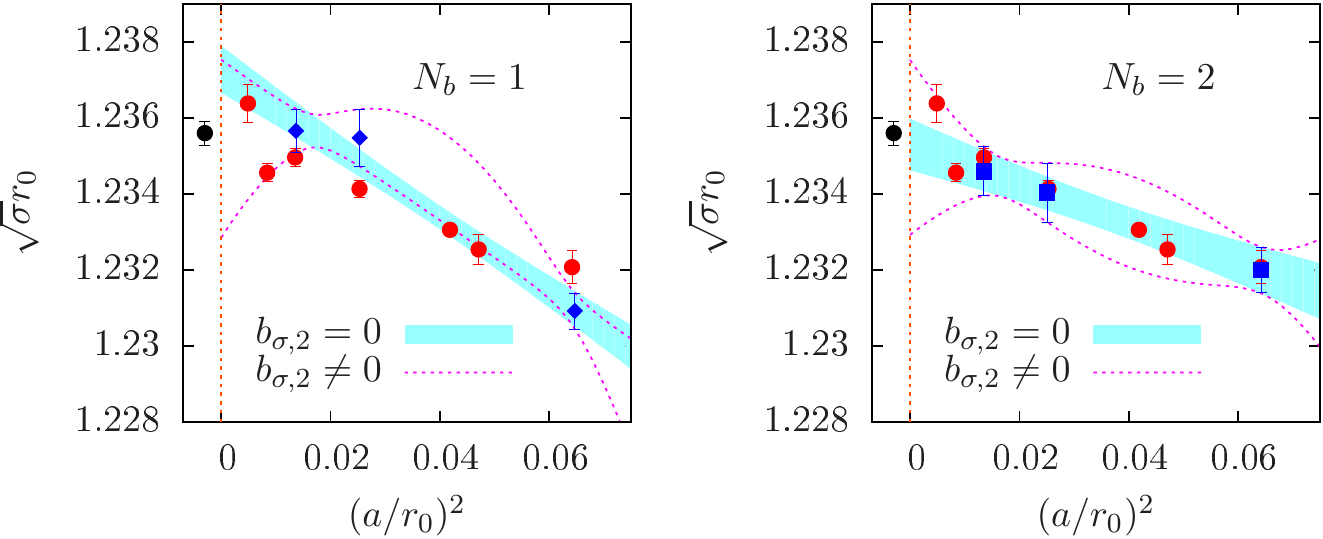}
 \caption{Continuum extrapolations of $\sqrt{\protect\vphantom{\sigma}\smash{\sigma_{(ii)}}}r_0$ from IPG for
 $\Nb=1$ (left) and $\Nb=2$ (right). For comparison we also show the results
 for $\SU(2)$ WPG from~\cite{Brandt:2017yzw} (red circles) and the associated
 continuum-extrapolated value (black circle).}
 \label{fig:sig_conti}
\end{figure}

To compare the results for $\sigma$ with the results for WPG with gauge group
$\SU(2)$ from~\cite{Brandt:2017yzw} we show the two sets of results in
figure~\ref{fig:sig_conti} for $\Nb=1$ (left) and $\Nb=2$ (right). For $\Nb=2$
we observe good agreement between the results from IPG and WPG, while some
slight differences are visible for $\Nb=1$. The latter could either be
fluctuations, remnants of uncontrolled systematic uncertainties connected to
the extraction of $\sigma$, or simply due to the different lattice artifacts
in the two theories. Eventually, we expect the results to agree in the continuum
limit. To test this, we have performed a continuum extrapolation of the
form\footnote{Due to the issues with the perturbative investigation of the
continuum limit for $\alpha\to1$ (see~\cite{Brandt:2016duy}) it is difficult to
obtain information about the powers of $a$ which contribute to
eq.~\eqref{eq:sig-conti} in IPG. However, leading corrections to the Wilson
action for $\Nb\to\infty$ at fixed $\alpha$ simply contain higher powers of
$(U+U^\dagger-2)$ in the action. These lead to corrections starting at
order $a^2$.}
\be
\label{eq:sig-conti}
\sqrt{\sigma} r_0 = \big( \sqrt{\sigma} r_0 \big)^\text{cont} + b_{\sigma,1}
\Big(\frac{a}{r_0}\Big)^2 + b_{\sigma,2} \Big(\frac{a}{r_0}\Big)^4 \,.
\ee
As in~\cite{Brandt:2017yzw} we perform two fits, either with $b_{\sigma,2}\neq0$
or with $b_{\sigma,2}=0$. The resulting extrapolations are also shown in
figure~\ref{fig:sig_conti} together with the extrapolation for WPG
from~\cite{Brandt:2017yzw} (black circle). We see that for $\Nb=2$ the continuum
extrapolations are all in very good agreement with the WPG result, even though
less precise, which can be expected since only three points are available for the
extrapolation. For $\Nb=1$ the extrapolation linear in $a^2$ overshoots the WPG result.
However, the data also indicate the importance of higher-order terms. Including the
$a^4$ term leads to an extrapolation which is fully consistent with the WPG
result, albeit with large uncertainties. For comparisons we will use the results
from the continuum extrapolation with $b_{\sigma,2}\neq0$. This extrapolation has
larger errors, but the central value agrees well with the continuum-extrapolated
WPG result in both cases.

\subsection{Results for the static potential}

\begin{figure}[t]
 \centering
\includegraphics[width=14cm]{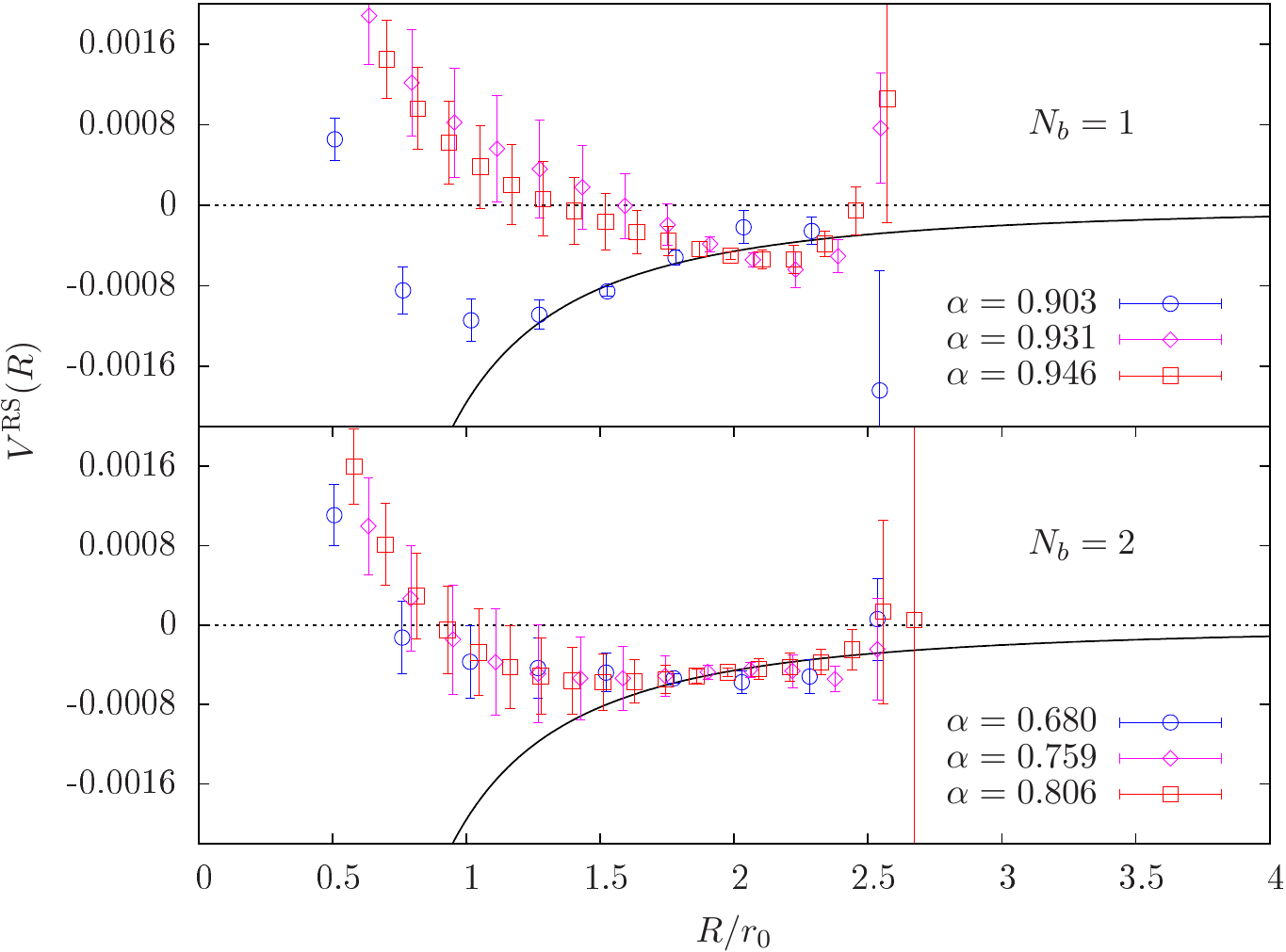}
 \caption{Results for the static potential in its rescaled form, see
eq.~\eqref{eq:resc-pot}, for different lattice spacings vs $R$ for SU(2) IPG
with $\Nb=1$ (top) and $\Nb=2$ (bottom). The black continuous line is the light-cone
potential obtained with the continuum-extrapolated string tension.}
 \label{fig:res_pot}
\end{figure}

We will now look at the results for the static potential itself. The results
for the potential, rescaled according to
\be
\label{eq:resc-pot}
V^\text{RS}(R) = \Big( \frac{V(R)-V_0}{\sqrt{\sigma}} - R\sqrt{\sigma} \Big)
\frac{R\sqrt{\sigma}}{\pi} + \frac{1}{24} \,,
\ee
are shown in figure~\ref{fig:res_pot}. As in~\cite{Brandt:2017yzw} the results
for the individual couplings have been rescaled using the string tension for
this value of $\alpha$, while the solid line is the rescaled version of the
leading-order EST prediction, the LC spectrum, with the continuum limit of the
string tension.

Results for the static potential in WPG were already presented in
figure~3 of Ref.~\cite{Brandt:2017yzw}. We do not reproduce that figure
here. The comparison with figure~\ref{fig:res_pot} shows one of the
main problems we are facing, namely the reduced accuracy of the
present study, which is mainly due to the less efficient algorithm
for IPG compared to the heat-bath algorithm used in the WPG simulations. This
leads to a reduction in the range of $R$ and reduces the precision in the
extraction of $\sigma$. Concerning the results for the potential itself, the
agreement with the WPG results in \cite[fig.~3]{Brandt:2017yzw} is eminent for $\Nb=2$ and also
visible for $\Nb=1$, although the lattice with $\alpha=0.903$
appears to be outside the scaling region.\footnote{This can be seen by
comparison to the data for $\beta=11.0$ in $\SU(3)$ gauge theory
in~\cite{Brandt:2017yzw}, where a similar behavior occurs.} As in WPG, the
corrections to the LC potential in IPG are positive and tend to become stronger when
approaching the continuum limit.

\begin{figure}[t]
 \centering
\includegraphics[width=14cm]{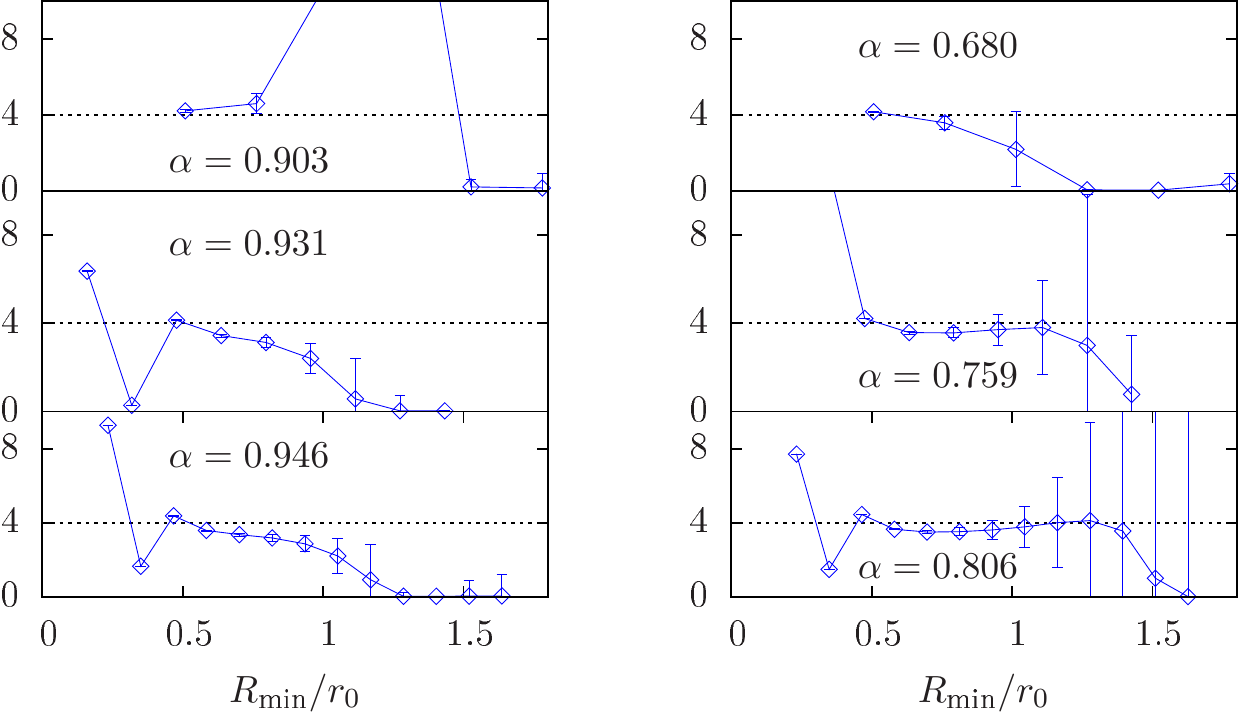}
 \caption{Results for the exponent $m$ plotted vs the minimal value of $R$
 included in the fit, $R_\text{min}$, for $\SU(2)$ IPG with $\Nb=1$ (left) and
 $\Nb=2$ (right). The horizontal line indicates the exponent of the leading
 correction to the EST.}
 \label{fig:m-exponent}
\end{figure}

The next step in the analysis is to check whether the leading-order correction
to the square-root formula in eq.~\eqref{eq:est-spec} is indeed of order $R^{-4}$.
To this end we fit the data to the form
\be
\label{eq:lead-coeff-fit}
V(R) = \sigma R \, \sqrt{ 1 - \frac{\pi(d-2)}{12\sigma R^{2}} }
 + \frac{\eta\sqrt\sigma}{\big(\sqrt{\sigma} R\big)^m} + V_0 \,,
\ee
where $\eta$, $m$, $\sigma$ and $V_0$ are fit parameters. If the predictions
of the EST are correct, we will obtain $m=4$. Otherwise we will find $m<4$
if the square-root formula in eq.~\eqref{eq:lead-coeff-fit} is incorrect,
or $m>4$ if the corrections start at higher order. We show the results for
$m$ vs $R_\text{min}$ in figure~\ref{fig:m-exponent}. As in $\SU(2)$ WPG,
cf.~\cite[fig.~4]{Brandt:2017yzw}, we typically observe a plateau around
$0.5\lesssim R/r_0 \lesssim 1.0$. For $\Nb=1$ the uncertainties are typically
larger for larger $R$-values, so that the plateau does not last as long as
for $\Nb=2$. The plateau value is typically around $m=3.6$.
This slight discrepancy with expectations has also been found in
WPG~\cite{Brandt:2017yzw} and indicates a possible mixing with other
correction terms. At finite $\Nc$ the EST will receive corrections
from virtual glueball exchange, for instance.

\subsection{Extraction of the boundary coefficient}

To make the comparison of the subleading properties of the potential more
quantitative, we will now extract the boundary coefficient \btt{}. As
in~\cite{Brandt:2017yzw}, we include higher-order terms in the fit formula
and fit the potential to the form
\be
\label{eq:boundary-fit}
V(R) = V^\text{EST}(R) + \frac{\gamma^{(1)}_{0}}{\sqrt{\sigma^5} R^6} + 
\frac{\gamma^{(2)}_{0}}{\sigma^3 R^7} + V_0 \,.
\ee
Specifically, we perform the following fits~\cite{Brandt:2017yzw}:
\begin{enumerate}
 \item[\textbf{A}] We use $\sigma$ and $V_0$ from method (ii) as input and
 fit with \btt,  $\gamma^{(1)}_{0}$ and $\gamma^{(2)}_{0}$ as free parameters.
 \item[\textbf{B}] We use $\sigma$, $V_0$  and \btt{} as free parameters and set
 $\gamma^{(1)}_{0}=0$ and $\gamma^{(2)}_{0}=0$.
 \item[\textbf{C}] We use $\sigma$, $V_0$,  \btt{} and $\gamma^{(1)}_{0}$ as free
 parameters and set $\gamma^{(2)}_{0}=0$.
 \item[\textbf{D}] We use $\sigma$, $V_0$, \btt{} and $\gamma^{(2)}_{0}$ as free
 parameters and set $\gamma^{(1)}_{0}=0$.
 \item[\textbf{E}] We use $\sigma$, $V_0$, $\gamma^{(1)}_{0}$ and $\gamma^{(2)}_{0}$
 as free parameters and set $\bt=0$.
\end{enumerate}
The fits are performed for several values of $R_\text{min}$, and we extract the final
result from the second smallest $R_\text{min}$ which provides a $\chi^2/\text{dof}<1.5$.
The quality of the agreement with the data is then indicated by the value of
$R_\text{min}$ (smaller values mean better agreement) in context with the number
of higher-order terms included in the fit. Fit \textbf{C}, for instance, should allow
for a smaller value of $R_\text{min}$ compared to fit \textbf{B} since the latter does
not contain higher-order correction terms. To check for the systematic uncertainty
associated with the fit interval we compare the result to the ones obtained with
$R_\text{min}\pm a$.

\begin{table}
\small
\centering
\begin{tabular}{@{\hspace{1mm}}c|ll|l|l@{\hspace{2.5mm}}l|c@{\hspace{2.5mm}}c@{\hspace{1mm}}}
 \hline
 \hline
 Fit & \multicolumn{1}{c}{$\sqrt{\sigma}r_0$} & \multicolumn{1}{c|}{$aV_0$} &
\multicolumn{1}{c|}{$\bt\cdot10^{2}$} &
\multicolumn{1}{c}{$\gamma^{(1)}_{0}\cdot10^{3}$} &
\multicolumn{1}{c|}{$\gamma^{(2)}_{0}\cdot10^{3}$} & $\chi^2/$dof &
$R_\text{min}/r_0$ \\
 \hline
 \hline
 \multicolumn{2}{l}{{ }$\alpha=0.903$} & & & & & & \\
 \textbf{A} & \multicolumn{1}{c}{---} & \multicolumn{1}{c|}{---} & 1.76(20)(235) &
 43(4)(79) & -27(4)(75) & 0.44 & 0.76 \\
 \textbf{B} & 1.2317(3)(5) & 0.2060(3)({ }5) & -1.73(4)(21) &
 \multicolumn{1}{c}{---} & \multicolumn{1}{c|}{---} & 0.59 & 0.76 \\
 \textbf{C} & 1.2314(5)(9) & 0.2064(6)(13) & -1.13(55)(449) & 2(2)(28) &
 \multicolumn{1}{c|}{---} & 0.61 & 0.76 \\
 \textbf{D} & 1.2314(5)(9) & 0.2064(6)(12) & -1.30(41)(323) &
 \multicolumn{1}{c}{---} & 1(2)(23) & 0.61 & 0.76 \\
 \textbf{E} & 1.2313(4)(6) & 0.2066(5)(8) & \multicolumn{1}{c|}{---} &
 20(10)(54) & -12(12)(62) & 0.56 & 0.76 \\
 \hline
 \multicolumn{2}{l}{{ }$\alpha=0.931$} & & & & & & \\
 \textbf{A} & \multicolumn{1}{c}{---} & \multicolumn{1}{c|}{---} & -4.61(194)(49)
 & -20(18)(7) & 10(10)(5) & 0.65 & 0.64 \\
 \textbf{B} & 1.2347(3)(4) & 0.1699(2)(3) & -2.46(10)(42) &
 \multicolumn{1}{c}{---} & \multicolumn{1}{c|}{---} & 0.47 & 0.96 \\
 \textbf{C} & 1.2349(3)(4) & 0.1697(2)(3) & -3.27(32)(128) & 4(2)(8) &
 \multicolumn{1}{c|}{---} & 0.37 & 0.80 \\
 \textbf{D} & 1.2349(3)(3) & 0.1697(2)(3) & -2.95(23)(95) &
 \multicolumn{1}{c}{---} & 3(1)(6) & 0.38 & 0.80 \\
 \textbf{E} & 1.2348(4)(5) & 0.1698(3)(4) & \multicolumn{1}{c|}{---} &
 73(71)(79) & -65(62)(99) & 0.41 & 0.96 \\
 \hline
 \multicolumn{2}{l}{{ }$\alpha=0.946$} & & & & & & \\
 \textbf{A} & \multicolumn{1}{c}{---} & \multicolumn{1}{c|}{---} & -3.56(114)(23)
 & -9(9)(4) & 4(5)(3) & 1.06 & 0.58 \\
 \textbf{B} & 1.2352(2)(3) & 0.1436(1)(2) & -2.48(7)(25) &
 \multicolumn{1}{c}{---} & \multicolumn{1}{c|}{---} & 0.47 & 0.94 \\
 \textbf{C} & 1.2352(2)(4) & 0.1436(1)(2) & -2.88(10)(44) & -2(1)(1) &
 \multicolumn{1}{c|}{---} & 0.39 & 0.70 \\
 \textbf{D} & 1.2352(2)(4) & 0.1436(1)(2) & -2.65(7)(36) &
 \multicolumn{1}{c}{---} & -2(1)(2) & 0.45 & 0.70 \\
 \textbf{E} & 1.2352(3)(3) & 0.1436(1)(2) & \multicolumn{1}{c|}{---} &
 61(61)(32) & -51(51)(38) & 0.36 & 0.94 \\
 \hline
 \hline
\end{tabular}
\caption{Results of the fits for the extraction of \btt{} for $\Nb=1$ IPG.}
\label{tab:b2-fits-n1}
\end{table}

\begin{table}
\small
\centering
\begin{tabular}{@{\hspace{1mm}}c|ll|l|l@{\hspace{2.5mm}}l|c@{\hspace{2.5mm}}c@{\hspace{1mm}}}
 \hline
 \hline
 Fit & \multicolumn{1}{c}{$\sqrt{\sigma}r_0$} & \multicolumn{1}{c|}{$aV_0$} &
\multicolumn{1}{c|}{$\bt\cdot10^{2}$} &
\multicolumn{1}{c}{$\gamma^{(1)}_{0}\cdot10^{3}$} &
\multicolumn{1}{c|}{$\gamma^{(2)}_{0}\cdot10^{3}$} & $\chi^2/$dof &
$R_\text{min}/r_0$ \\
 \hline
 \hline
 \multicolumn{2}{l}{{ }$\alpha=0.680$} & & & & & & \\
 \textbf{A} & \multicolumn{1}{c}{---} & \multicolumn{1}{c|}{---} & -4.57(460)(288) &
 -31(58)(93) & 18(37)(88) & 0.38 & 0.76 \\
 \textbf{B} & 1.2317(2)(5) & 0.2114(2)(5) & -1.73(3)(17) &
 \multicolumn{1}{c}{---} & \multicolumn{1}{c|}{---} & 0.29 & 0.76 \\
 \textbf{C} & 1.2320(4)(4) & 0.2111(4)(6) & -2.17(39)(183) & -2(2)(12) &
 \multicolumn{1}{c|}{---} & 0.25 & 0.76 \\
 \textbf{D} & 1.2320(4)(4) & 0.2111(4)(5) & -2.05({ }30)(134) &
 \multicolumn{1}{c}{---} & -1(1)(10) & 0.25 & 0.76 \\
 \textbf{E} & 1.2317(3)(7) & 0.2115(4)(8) & \multicolumn{1}{c|}{---} &
 29(29)(45) & -20(20)(52) & 0.34 & 0.76 \\
 \hline
 \multicolumn{2}{l}{{ }$\alpha=0.759$} & & & & & & \\
 \textbf{A} & \multicolumn{1}{c}{---} & \multicolumn{1}{c|}{---} & -0.11(252)(119) &
 34(30)(34) & -24(18)(30) & 0.26 & 0.79 \\
 \textbf{B} & 1.2342(3)(2) & 0.1723(2)(2) & -2.25(10)(16) &
 \multicolumn{1}{c}{---} & \multicolumn{1}{c|}{---} & 0.18 & 0.95 \\
 \textbf{C} & 1.2343(3)(1) & 0.1722(2)(1) & -2.65(30)(22) & -2(1)(1) &
 \multicolumn{1}{c|}{---} & 0.17 & 0.79 \\
 \textbf{D} & 1.2343(3)(2) & 0.1722(2)(1) & -2.50(22)(20) &
 \multicolumn{1}{c}{---} & -2(1)(1) & 0.17 & 0.79 \\
 \textbf{E} & 1.2339(3)(5) & 0.1726(2)(4) & \multicolumn{1}{c|}{---} &
 34(34)(13) & -24(24)(14) & 0.34 & 0.76 \\
 \hline
 \multicolumn{2}{l}{{ }$\alpha=0.806$} & & & & & & \\
 \textbf{A} & \multicolumn{1}{c}{---} & \multicolumn{1}{c|}{---} & -1.16(267)({ }99) &
 15(28)(21) & -10(16)(16) & 0.50 & 0.70 \\
 \textbf{B} & 1.2346(2)(3) & 0.1445(1)(2) & -2.21(4)(15) &
 \multicolumn{1}{c}{---} & \multicolumn{1}{c|}{---} & 0.19 & 0.81 \\
 \textbf{C} & 1.2349(2)(3) & 0.1443(1)(2) & -2.67(12)(32) & -2(1)(1) &
 \multicolumn{1}{c|}{---} & 0.02 & 0.70 \\
 \textbf{D} & 1.2348(2)(3) & 0.1444(1)(2) & -2.50(9)(26) &
 \multicolumn{1}{c}{---} & -1.1(2)(7) & 0.03 & 0.70 \\
 \textbf{E} & 1.2345(3)(4) & 0.1446(1)(2) & \multicolumn{1}{c|}{---} &
 57(37)(101) & -26(26)(11) & 0.08 & 0.81 \\
 \hline
 \hline
\end{tabular}
\caption{Results of the fits for the extraction of \btt{} for $\Nb=2$ IPG.}
\label{tab:b2-fits-n2}
\end{table}

We list the results of the different fits for $\Nb=1$ and 2 in
tables~\ref{tab:b2-fits-n1} and~\ref{tab:b2-fits-n2}, respectively. Comparing
the results to those of~\cite[tab. 5]{Brandt:2017yzw}, in particular for
$\beta=5.0$, 7.5 and 10.0, we see that both the possible fit ranges and the
resulting parameters are very similar. The only exception is fit \textbf{A},
for which $\sigma$ and $V_0$ have been taken over from
section~\ref{sec:sim-points}. Since the extraction of $\sigma$ and $V_0$ was
not as accurate as in~\cite{Brandt:2017yzw} it is reasonable that
this is the main reason for deviations in this particular fit. We can thus
follow the discussion of~\cite{Brandt:2017yzw} and conclude that fits \textbf{B},
\textbf{C} and \textbf{D} can be used in the following analysis. Fit \textbf{E} typically
requires larger values for $R_\text{min}$ compared to fits \textbf{C} and \textbf{D},
even though it also includes two higher-order terms.
Thus we conclude that the agreement with the data is worse for fit \textbf{E}
than for the other fits. In any case, fit \textbf{E} only serves as a
check that, given the data, \btt{} does not vanish. Compared
to~\cite{Brandt:2017yzw}, where fit \textbf{E} clearly showed less agreement with
the data, direct conclusions are more difficult here since the IPG data are
less precise than those of WPG.

\begin{figure}[t]
 \centering
 \includegraphics[]{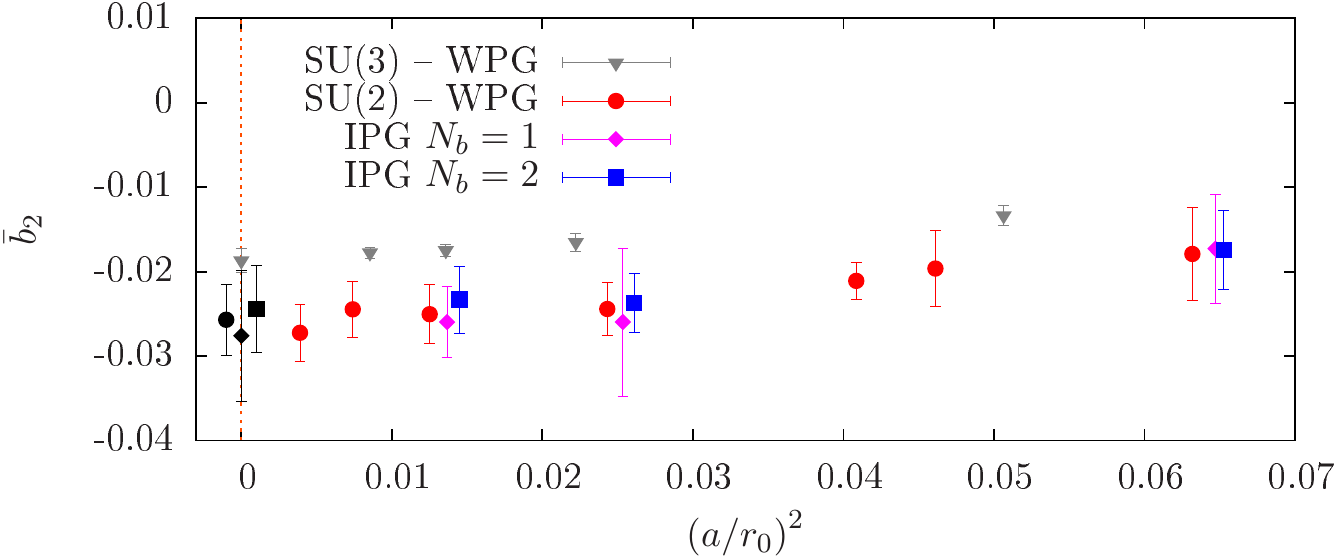}
 \caption{Results for \btt{} from table~\ref{tab:b2-results} and for $\SU(2)$
 WPG from~\cite{Brandt:2017yzw} vs the squared lattice spacing in units
 of $r_0$. Also shown are the continuum results from eq.~\eqref{eq:b2-conti-res-ipg}
 and the $\SU(2)$ result from~\cite[eq.~(6.2)]{Brandt:2017yzw}. The results for
 $\SU(2)$ WPG and for $\Nb=2$ have been slightly shifted to the left and to the
 right, respectively, to enable the visibility of the different sets of points.}
 \label{fig:br_vs_r0}
\end{figure}

As in~\cite{Brandt:2017yzw}, we determine the final results for \btt{} on the
individual lattice spacings via the weighted average over the results from fits
\textbf{B} to \textbf{D}, where the weight is given by the uncertainties. To estimate
the systematic uncertainty due to the choice of $R_\text{min}$ we repeat the
procedure with $R_\text{min}\pm a$ and use the maximal deviation. The results are
tabulated in table~\ref{tab:b2-results}. Comparing once more to the $\SU(2)$
results of~\cite{Brandt:2017yzw}, which we list for comparison in
table~\ref{tab:b2-results} as well, we see that the results are similar in
magnitude and have similar uncertainties. This is particularly true for $\Nb=2$.
For $\Nb=1$ the systematic uncertainties are somewhat larger, but the overall
agreement is still good. We show the results for \btt{} in comparison to the
results of~\cite{Brandt:2017yzw} in figure~\ref{fig:br_vs_r0}. One can clearly
see the similar behavior in the approach to the continuum and the good agreement
between the results from the different simulations. In particular, the results
are significantly different from the ones for $\SU(3)$ gauge theory, showing
the discriminating power of the results.

\begin{table}[t]
\small
\centering
\begin{tabular}{cc|cc|cc}
 \hline
 \hline
 \multicolumn{2}{c|}{WPG} &
 \multicolumn{2}{c|}{IPG $\Nb=1$} & \multicolumn{2}{c}{IPG $\Nb=2$} \\
 $\beta$ & \btt & $\alpha$ & \btt & $\alpha$ & \btt \\
 \hline
 \hline
 5.0 & -0.0179({ }5)(50)(23) & 0.903 & -0.0173({ }4)(59)(24) & 0.680 & -0.0174({ }4)(43)(19) \\
 7.5 & -0.0244(11)(25)(16) & 0.931 & -0.0260(14)(68)(54) & 0.759 & -0.0237(16)(28)(14) \\
 10.0 & -0.0251({ }5)(27)(22) & 0.946 & -0.0260({ }7)(29)(31) & 0.806 & -0.0233({ }6)(35)(19) \\
 \hline
 \hline
\end{tabular}
\caption{Final results for \btt{} in WPG~\cite{Brandt:2017yzw} and IPG with
$\Nb=1$ and 2 for the individual couplings. The first error is the statistical
uncertainty, the second the systematic one due to the unknown higher-order correction
terms, estimated by computing the maximal deviations of the results from fits
\textbf{B} to \textbf{D}, and the third is the systematic one associated with the choice
of $R_\text{min}$.}
\label{tab:b2-results}
\end{table}

\subsection{Continuum limit}

\begin{figure}[t]
 \centering
\includegraphics[]{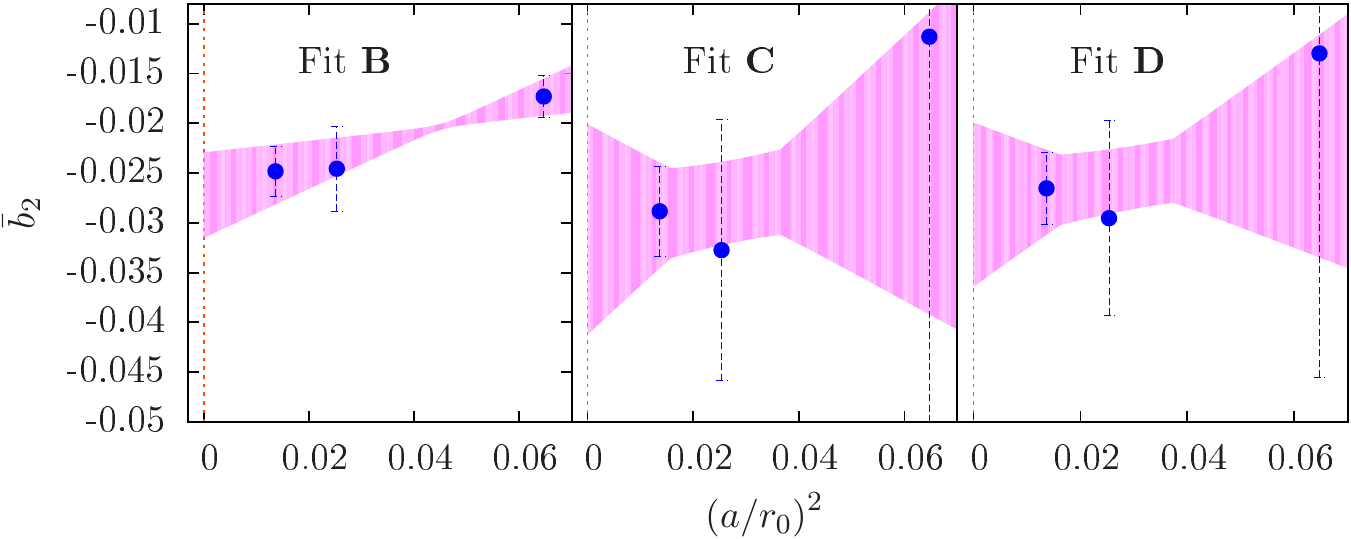} \\[2mm]
\includegraphics[]{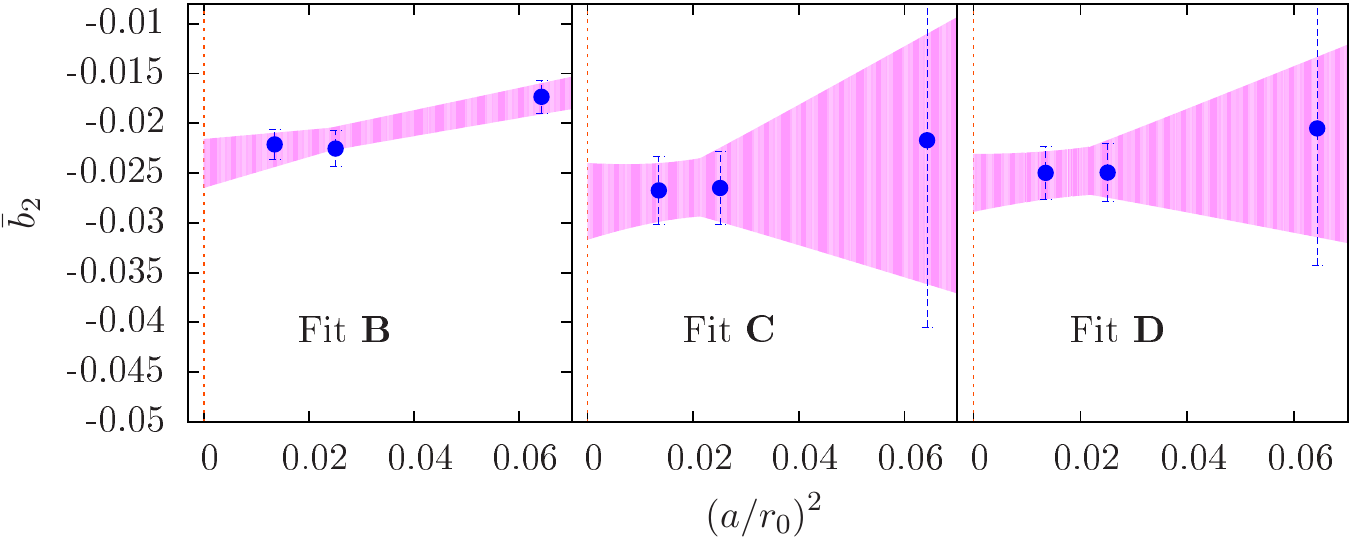} 
 \caption{Results for the linear continuum extrapolation, fit (2), for the
results for \btt{} obtained from fits \textbf{B}, \textbf{C} and \textbf{D} (from left to
right) for $\SU(2)$ IPG with $\Nb=1$ (top) and $\Nb=2$ (bottom).}
 \label{fig:br_vs_r0_fit}
\end{figure}

Finally, we perform the continuum extrapolation of the boundary coefficient
\btt. To this end we parameterize it as~\cite{Brandt:2017yzw}
\be
\label{eq:b2-conti}
\bt = \big( \bt \big)^\text{cont} + b_{\bt,1} \Big(\frac{a}{r_0}\Big)^2 + b_{\bt,2}
\Big(\frac{a}{r_0}\Big)^4
\ee
and perform two different types of fits:
\begin{enumerate}
 \item[(1)] We fit including all terms in eq.~\eqref{eq:b2-conti}.
 \item[(2)] We fit with setting $b_{\bt,2}=0$.
\end{enumerate}
Note that fit (1) is a parameterization of the results rather than a fit.
In~\cite{Brandt:2017yzw} a third fit has been performed, which included
only the data for the finest lattice spacings. Such a fit does not make sense
here due to the limited number of available couplings. To estimate the
propagation of systematic uncertainties we follow the strategy
of~\cite{Brandt:2017yzw} and perform the fits (1) and (2) for the results
from fits \textbf{B} to \textbf{D} and the fits with a minimal $R$-value of
$R_\text{min}\pm a$ individually. The final results have been extracted using
a weighted average of the results from fits \textbf{B} to \textbf{D}, once more
with the individual uncertainties as weights.
As before, the individual systematic uncertainties are computed
from the maximal deviations to the
different fits. The curves for fit (2) with the main value of
$R_\text{min}$ are shown in figure~\ref{fig:br_vs_r0_fit}.

\begin{table}
\small
\centering
\begin{tabular}{cc|ccc}
 \hline
 \hline
 $\Nb$ & Fit & (1) & (2) \\
 \hline
 1 & $\big( \bt \big)^\text{cont}$ & -0.0223(37)(40)(44) & -0.0276({ }9)(31)(47) \\
 2 & $\big( \bt \big)^\text{cont}$ & -0.0214(26)(50)(29) & -0.0244({ }6)(34)(24) \\
 \hline
 \hline
\end{tabular}
\caption{Results for $\big( \bt \big)^\text{cont}$ from fits (1) and (2) (see text).
The first error is the statistical uncertainty, the second the one
associated with the unknown higher-order correction terms in the potential and
the third the one due to the choice of $R_\text{min}$.}
\label{tab:b2_conti_fitpar}
\end{table}

The continuum results for \btt{} from fits (1) and (2) are given in
table~\ref{tab:b2_conti_fitpar}. We use the results from fit (1)
to estimate the systematic uncertainty associated with the continuum
limit. The final results, which we take from fit (2), are thus given
by
\be
\label{eq:b2-conti-res-ipg}
\big( \bt \big)^\text{cont} = \left\{ \begin{array}{ll}
-0.0276(9)(31)(47)(53) & \quad \text{for} \:\:\Nb=1 \,, \\[2mm]
-0.0244(6)(34)(24)(30) & \quad \text{for} \:\:\Nb=2 \,.
\end{array} \right.
\ee
Those can be compared to the final continuum estimate for $\SU(2)$ WPG
from~\cite{Brandt:2017yzw},
\be
\label{eq:b2-conti-res-wpg}
\big( \bt \big)^\text{cont} =
-0.0257\,(3)(38)(17)(3) \,.
\ee
In eqs.~\eqref{eq:b2-conti-res-ipg} and~\eqref{eq:b2-conti-res-wpg} the first
error is purely statistical, the second is the systematic uncertainty due to
the unknown higher-order terms in the potential, the third is the one
associated with the particular choice for $R_\text{min}$ and the fourth
is the systematic uncertainty due to the continuum extrapolation.
We see that the results from eqs.~\eqref{eq:b2-conti-res-ipg}
and~\eqref{eq:b2-conti-res-wpg} agree well within uncertainties. In addition,
the sizes of the individual uncertainties are similar, except for the continuum
extrapolation, which, however, is expected since in the IPG analysis ensembles at fewer
lattice spacings are available.

We briefly summarize the findings of this section. We have repeated the
analysis of the potential in~\cite{Brandt:2017yzw} for IPG and found that in
every individual step the results agree extremely well with WPG, for both 
$\Nb=1$ and $2$. The whole analysis indicates
that the fine structure of the potential in the continuum is indeed identical
in IPG and WPG. Hence we can conclude that, at least for the potential,
both theories lead to the same continuum limit (up to the accuracy of this
study). This is also reflected in the significant difference to the results
for $\SU(3)$ WPG, which shows the discriminating power of this comparison.

\section{The finite-temperature phase transition}
\label{sec:finiteT}

So far we have compared properties of IPG and WPG
for observables at vanishing temperature and found good
agreement. We will now show that the agreement prevails for
thermodynamic observables. In particular, we will consider
the deconfinement transition temperature $\Tc$ and the ratio of critical
exponents $\gamma/\nu$. The latter can be regarded as a measure of the
universality class of the transition (we expect a phase transition of 2nd order).
The fundamental lattice observable that can be used to investigate
the deconfinement transition is the absolute value of the Polyakov loop,
$\ev{|L|}$, the order parameter associated with the breaking of center
symmetry. In particular, we choose a setup which is similar to that
in~\cite{Engels:1996dz} so that we can directly compare
to this study.

\subsection{Simulation parameters and results for the Polyakov loop}

\begin{table}[t]
  \centering
  \small
  \begin{tabular}{l|cl|c|cc}
    \hline
    \hline
    Theory & $N_t$ & \multicolumn1{c|}{volumes} & couplings & \# meas & $N_\text{sw}$ \\
    \hline
    \hline
    WPG & 4 & $32^2$, $48^2$, $64^2$, $96^2$ & 5.00-7.50 & 100k & $\diagup$ \\
    IPG $\Nb=1$ & 4 & $32^2$, $48^2$, $64^2$, $96^2$ & 0.900-0.945 & 100k-400k &
    100 \\
    IPG $\Nb=2$ & 4 & $32^2$, $48^2$, $64^2$, $96^2$ & 0.650-0.830 & 100k-200k &
    100 \\
    \hline
    WPG & 6 & $48^2$, $72^2$, $96^2$, $144^2$ & 8.00-11.50 & 100k & $\diagup$ \\
    IPG $\Nb=1$ & 6 & $48^2$, $72^2$, $96^2$, $144^2$ & 0.930-0.950 & 400k &
    100 \\
    IPG $\Nb=2$ & 6 & $48^2$, $72^2$, $96^2$, $144^2$ & 0.750-0.850 & $\sim$400k &
    100 \\
    \hline
    \hline
  \end{tabular}
  \caption{Simulation parameters of the finite-temperature runs in 3d $\SU(2)$ IPG
    and WPG theories. For each volume the simulations have been done on the same
    values of the coupling. For WPG and $\Nt=4$ the distance between two $\beta$-values
    was taken to be $0.01$ in the region around $T_c(\infty)$, while for $\Nt=6$
    we used a distance of $0.05$. For IPG theory with $\Nb=1$, the distance between
    two $\alpha$-values was taken to be $0.0001$ for $\Nt=4$ and $0.00025$ for $\Nt=6$,
    while for $\Nb=2$ we used $0.001$ for both $\Nt=4$ and 6. Away from $T_c(\infty)$ we have
    simulated less frequently.  For $\Nt=4$ the number of measurements varies between
    the different volumes, and around $T_c(\infty)$ we have increased the number of
    measurements with the volume.}
  \label{tab:simpoints_finiteT}
\end{table}

As before, we perform simulations in IPG using $\Nb=1$ and 2. In addition, we
also simulate in WPG for a direct comparison of the results. To test the approach
to the continuum limit we use two different temporal extents, $\Nt=4$ and 6, for
which we vary the temperature $T=1/(a\Nt)$ by varying the lattice spacing $a$ via
the lattice couplings $\alpha$ and $\beta$. For scale setting we use the results
of section~\ref{sec:matching}. The resulting simulation parameters are listed
in table~\ref{tab:simpoints_finiteT}. For $\Nt=6$ the volumes have been chosen
to match the volumes for $\Nt=4$ in physical units.

The main observables are the absolute value of the Polyakov loop
\be
\label{eq:poly-loop}
\vert L \vert = \frac{1}{V} \sum_{\vec{x}} \Big| \Tr \prod_{n_0=1}^{N_t}
U_0(n_0\,a,\vec{x}) \Big|
\ee
and its susceptibility
\be
\label{eq:poly-susc}
\chi_L = V \left( \left< \vert L \vert^2 \right> - \left< \vert L \vert \right>^2
\right) .
\ee
For each simulation point we have performed at least 100,000 measurements and
increased the number of measurements to about 400,000 in the vicinity
of $\Tc$, where the autocorrelations increase due to the approach of a second-order
critical point.

\begin{figure}[t]
  \centering
  \begin{tabular}{r@{\hspace{7mm}}r}
    \includegraphics[scale=.89]{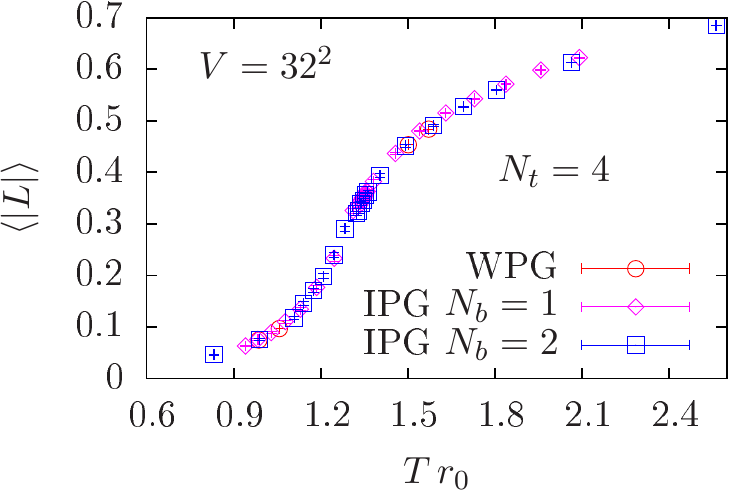}
    & \includegraphics[scale=.89]{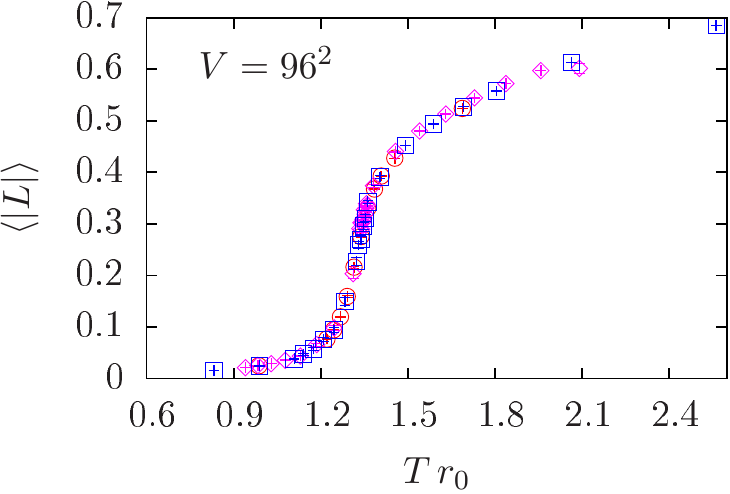}
    \\[3mm]
    \includegraphics[scale=.89]{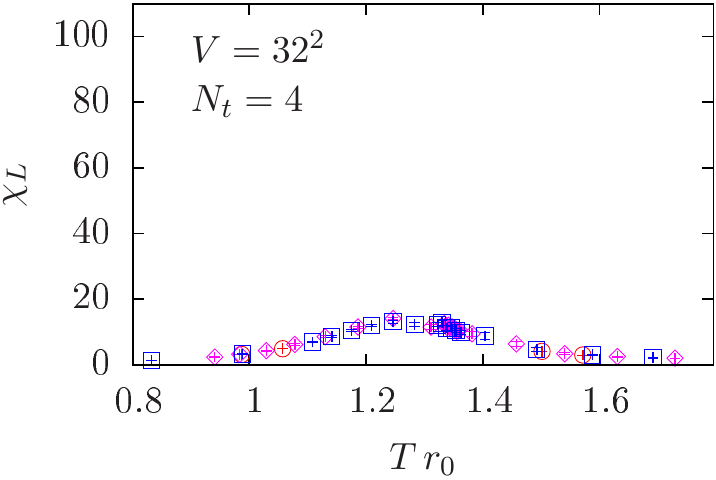}
    & \includegraphics[scale=.89]{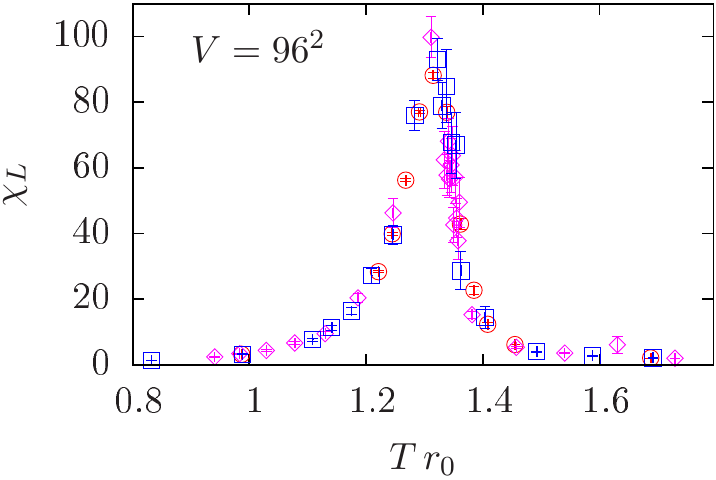}
  \end{tabular}
  \caption{Results for the absolute value of the Polyakov loop $\ev{|L|}$
    (top) and its susceptibility $\chi_L$ (bottom) at $N_t=4$ with volumes
    $32^2$ (left) and $96^2$ (right). The legend is the same for all
    plots.}
  \label{fig:polyakov_loops-nt4}
\end{figure}

\begin{figure}[t]
  \centering
  \begin{tabular}{r@{\hspace{7mm}}r}
    \includegraphics[scale=.89]{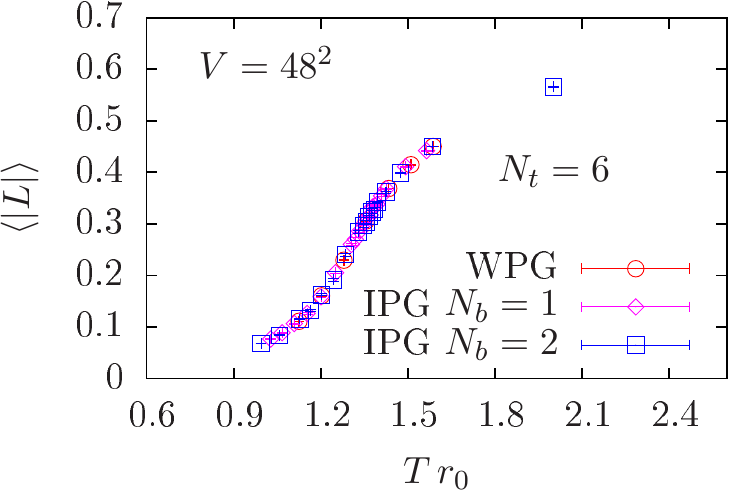}
    & \includegraphics[scale=.89]{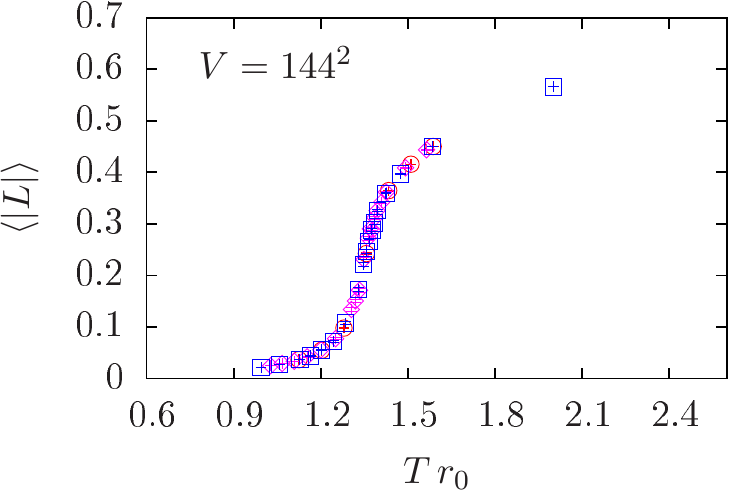}
    \\[3mm]
    \includegraphics[scale=.89]{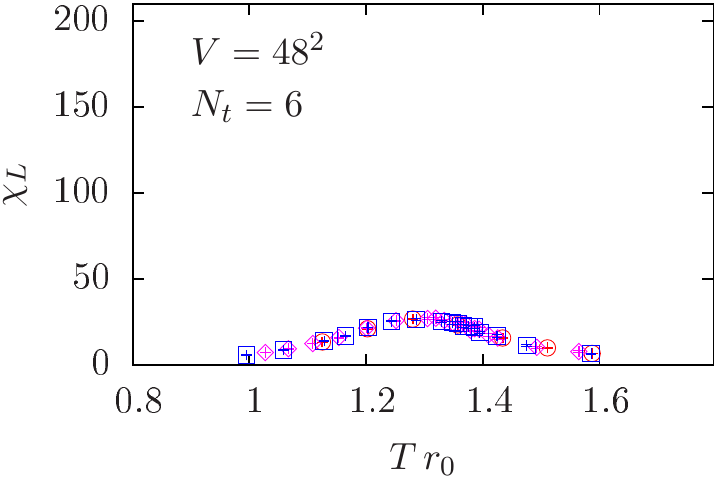}
    & \includegraphics[scale=.89]{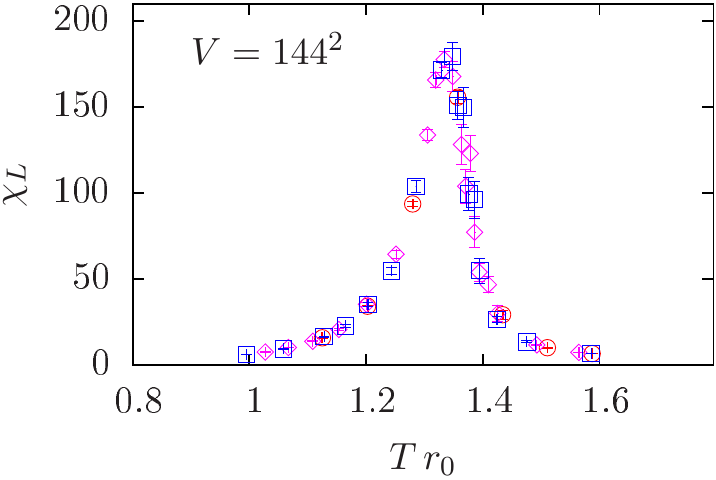}
  \end{tabular}
  \caption{Results for the absolute value of the Polyakov loop $\ev{|L|}$
    (top) and its susceptibility $\chi_L$ (bottom) at $N_t=6$ with volumes
    $32^2$ (left) and $96^2$ (right). The legend is the same for all
    plots.}
  \label{fig:polyakov_loops-nt6}
\end{figure}

The results for $\ev{|L|}$ and $\chi_L$ vs the temperature in units of
the Sommer parameter are shown in
figures~\ref{fig:polyakov_loops-nt4} and~\ref{fig:polyakov_loops-nt6} for
$\Nt=4$ and 6, respectively. We always show the two extremal cases of
smallest and largest available volume. The plots show the remarkable
agreement between the results of WPG and IPG with $\Nb=1$ and 2. This
already indicates the similarity between the corresponding phase transitions. In
particular, the volume scaling is equivalent in the different cases so
that the universality classes can be expected to be equivalent as well.
In the following we will investigate this expectation more quantitatively.

\subsection{The transition temperature}

We define the critical temperature $T_c$ through the peak
of the Polyakov-loop susceptibility. We determine it by fitting a
Gaussian to the points in the vicinity of the peak. This
definition assumes a Gaussian form of the susceptibility peak, which
will, generically, not be the case. To account for the associated
systematic uncertainty we use the distance between the two points
surrounding the maximum as a conservative error estimate for $\Tc$.
We have also checked that the systematic uncertainty associated with
the number of points (symmetrically distributed around the maximum)
entering the fit is much smaller than this estimate for the systematic
uncertainty of the results. The results for $T_c$ in units of $r_0$
are listed in table~\ref{tab:trans-temp} for the different values of
$\Nt$ and the different volumes. We have not listed the other
fit parameters, such as the width of the Gaussian, since they are not
relevant for the following analysis.

\begin{table}[t]
  \centering
  \small
  \begin{tabular}{cc|cccc|ccl}
    \hline
    \hline
    $\Nt$ & Theory & \multicolumn{4}{c|}{$T_cr_0$} & \\
    \hline
    \hline
    & & $32^2$ & $48^2$ & $64^2$ & $96^2$ & $T_cr_0(\infty)$ & $\gamma/\nu$ & \\
    \hline
    4 & WPG & 1.27(2) & 1.28(1) & 1.30(1) & 1.31(2) & 1.343(7) & 1.70(4) & \cite{Engels:1996dz} \\
    & IPG $\Nb=1$ & 1.26(5) & 1.28(3) & 1.30(5) & 1.30(5) & 1.33(10) & 1.33(61) & \\
    & IPG $\Nb=2$ & 1.27(2) & 1.28(4) & 1.30(2) & 1.31(3) & 1.34(8) & 1.48(48) & \\
    \hline
    & & $48^2$ & $72^2$ & $96^2$ & $144^2$ & & & \\
    \hline
    6 & WPG & 1.29(7) & 1.31(5) & 1.32(4) & 1.34(6) & 1.365(7) & 1.68(7) & \cite{Engels:1996dz} \\
    & IPG $\Nb=1$ & 1.29(4) & 1.30(3) & 1.31(1) & 1.34(1) & 1.39(3) & 1.81(9) & \\
    & IPG $\Nb=2$ & 1.29(4) & 1.30(5) & 1.31(3) & 1.33(2) & 1.36(6) & 1.58(29) & \\
    \hline
    \hline
  \end{tabular}
  \caption{Results for the transition temperatures on the
    different volumes together with the infinite-volume extrapolations
    and the ratio $\gamma/\nu$ of critical exponents (see text). The
    results of the Bielefeld group~\cite{Engels:1996dz} are quoted as
    the results for WPG for comparison.}
  \label{tab:trans-temp}
\end{table}

\begin{figure}[t]
  \centering
  \includegraphics[]{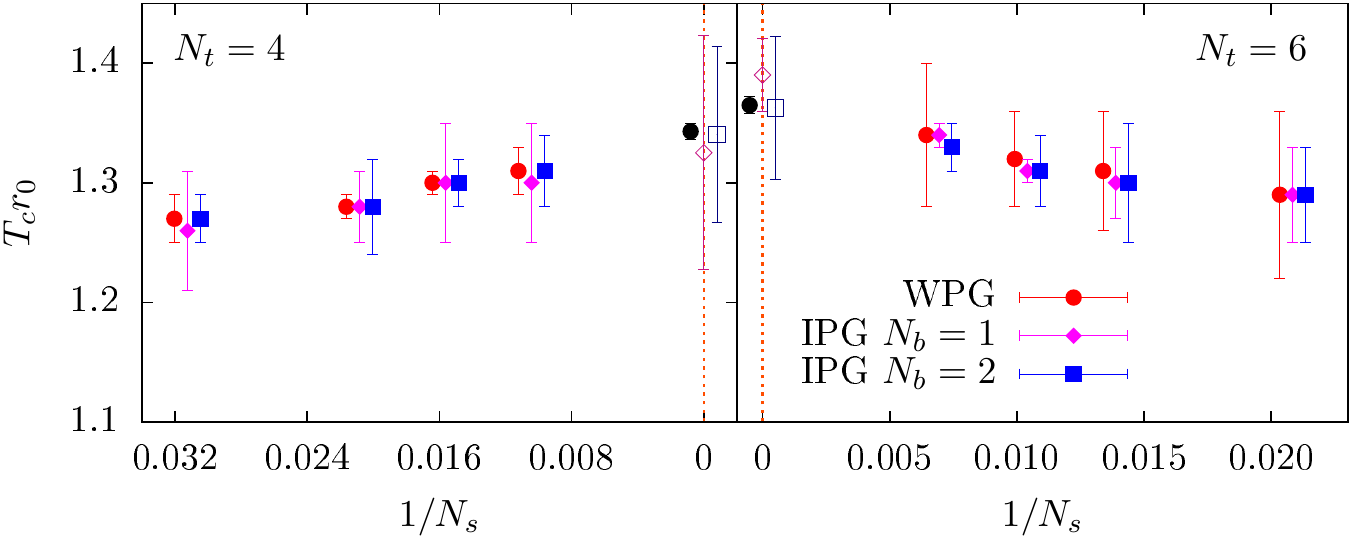}
  \caption{Results for the deconfinement transition temperatures $T_cr_0$
    vs the inverse box length $1/N_s$ for $\Nt=4$ (left) and $\Nt=6$
    (right). The data have been slightly displaced to improve the visibility
    of the different sets of points. The black data point at $1/N_s=0$ is
    the result in the infinite-volume limit from the Bielefeld
    group~\cite{Engels:1996dz}, while the magenta and
    blue open symbols are the results for $T_c(\infty)$ from IPG with
    $\Nb=1$ and 2, respectively.}
  \label{fig:tc-scaling}
\end{figure}

As expected from the results of the previous section, the results
for WPG and IPG with $\Nb=1$ and $\Nb=2$ agree well within
uncertainties. For $\Nt=6$ the uncertainties for WPG are larger due to
the larger separation of simulation points. The results can be compared
to the results of~\cite{Engels:1996dz} in the $V\to\infty$ limit.
To convert to units in terms of $r_0$ we use the results for
$\beta_c(\infty)$ (given in~\cite[table 1]{Engels:1996dz}) and convert
them to $T_cr_0(\infty)$ using the interpolation for $r_0(\beta)$
from section~\ref{sec:sim_par}. The results are also listed in
table~\ref{tab:trans-temp}, where the uncertainties include the
uncertainties for $\beta_c(\infty)$ and for $r_0/a$.
To be able to directly compare the results we
need to perform a $V\to\infty$ extrapolation of our data for $T_c$.
For $N_s\to\infty$, $T_c$ is expected to obey the scaling behavior
\be
\label{eq:tc-scaling}
T_cr_0(N_s) = h N_s^{-1/\nu} + T_cr_0(\infty) \,,
\ee
where $\nu$ is the associated critical exponent. The Potts model
associated with the $\SU(2)$ transition yields $\nu=1$~\cite{Wu:1982ra},
i.e., a linear behavior of $T_c$ with $1/N_s$, which has been found to be
in good agreement with the numerical data for $T_c$~\cite{Liddle:2008kk}.
Figure~\ref{fig:tc-scaling} is clearly in agreement with
eq.~\eqref{eq:tc-scaling}, even though the uncertainties are too large to
draw any final conclusions.
Assuming that the three largest volumes are already in the scaling region,
we perform a linear extrapolation to $N_s\to\infty$. The results are also
listed in table~\ref{tab:trans-temp}. We show the
results vs the inverse box length $1/N_s$ (the spatial volume is given
as $N_s^2$ in table~\ref{tab:trans-temp}) in
figure~\ref{fig:tc-scaling}. The plot indicates good agreement
between IPG, with both $\Nb=1$ and 2, and WPG at finite and infinite
volume. Note that our uncertainties are very conservative estimates for the
systematic uncertainties associated with the fits for the extraction
of $T_c$ and thus could be overestimated. These large uncertainties
also propagate to $T_c(\infty)$.

\subsection{Order and universality class of the transition}

Owing to the scaling of $T_c$ discussed above, we expect the 
transitions in IPG and WPG to be in the same universality class. Moreover, as
already discussed in section~\ref{sec:theory}, center symmetry
is a good symmetry for both actions so that the deconfinement
transition is accompanied by center-symmetry breaking. Another
strong indication for the transitions being in the same
universality class comes from the similar volume scaling of the
susceptibility peaks of the Polyakov loop (cf.\
figures~\ref{fig:polyakov_loops-nt4}
and~\ref{fig:polyakov_loops-nt6}). We will now test whether these
expectations are correct and both transitions are indeed in the 2d
Ising universality class, which is known to be the case at least
for 3d $\SU(2)$ WPG theory~\cite{Engels:1996dz}.

There are several types of analyses one can employ to determine
the critical exponents which distinguish the different universality
classes. Here we follow the strategy of~\cite{Engels:1996dz} and use
the $\chi^2$-method~\cite{Engels:1992ke}. The starting point is the
finite-size scaling formula for the susceptibility of the Polyakov
loop expanded around the critical point,
\begin{align}
  \label{eq:fss-scaling}
  \chi_L = N_s^{\gamma/\nu} \Big( c_0 + \Big( c_1 + c_2 N_s^{-\gamma_i}
  \Big) \tau N_s^{1/\nu} + c_3 N_s^{-\gamma_i} \Big) .
\end{align}
Here, $c_{0,1,2}$ are unknown coefficients, $\gamma_i$ is an unknown
exponent,
\begin{align}
  \label{eq:reduced_temp}
  \tau = \frac{T_c(\infty) - T}{T_c(\infty)}
\end{align}
is the reduced temperature with respect to the critical temperature in
the thermodynamic limit $T_c(\infty)$, $N_s$ is the spatial extent of
the lattice, and $\gamma/\nu$ is the desired ratio of critical exponents.
Exactly at $T_c(\infty)$, eq.~\eqref{eq:fss-scaling} reduces to
\begin{align}
  \label{eq:fss-scaling_attc}
  \chi_L = N_s^{\gamma/\nu} \left( c_0 + c_3 N_s^{-\gamma_i} \right) .
\end{align}
At large $N_s$ the second term (proportional to $c_3$) is a correction
and can be neglected.  We thus arrive at the scaling relation
\begin{align}
  \label{eq:fss-scaling_fin}
  \ln(\chi_L) = C + \frac{\gamma}{\nu} \ln(N_s) \,.
\end{align}
This scaling law can be tested by fitting the data at a given coupling
to \eqref{eq:fss-scaling_fin}. Since at $T\neq T_c(\infty)$ there are
corrections of the form given in eq.~\eqref{eq:fss-scaling}, we will
get the best fit when the coupling is equal (or very close) to the
critical coupling. The critical temperature $T_c(\infty)$ can thus be
extracted by looking at $\chi^2/$dof, in principle.

The main problem of this analysis is the finite resolution of simulation
points in the region around $T_c$ and the different accuracy of the results
for the susceptibility.  For WPG theory this problem can be overcome by
the use of the multi-histogram method~\cite{Ferrenberg:1988yz}, which can
be used for a well-controlled interpolation between simulation points. In
addition, the method leads to enhanced and balanced statistics for all
simulation points. In IPG theory this method cannot be applied since
$\alpha$, unlike $\beta$, does not appear as a simple prefactor in front
of an observable (the average plaquette for WPG) in the action. For IPG
this leads to the problem that $\chi^2/$dof fluctuates strongly and cannot
be used as a conclusive indicator for $T_c$ as 
in~\cite{Engels:1996dz}. Instead, we will compare to the results for
$\gamma/\nu$ obtained for the individual temperatures by fitting to
eq.~\eqref{eq:fss-scaling_fin} and making use of the results for $T_c(\infty)$
from table~\ref{tab:trans-temp}. In this way we obtain a number of possible
results for $\gamma/\nu$ in the region of $T_c(\infty)$. To be conservative,
we use the full spread of results as the uncertainty interval which encloses
the final result for $\gamma/\nu$ and define the central value of our final
result to be the midpoint of this interval. These results are also listed in
table~\ref{tab:trans-temp}. Unfortunately, the uncertainties in IPG are too
large to draw definite conclusions from the comparison. Nonetheless, the
results are in agreement with those of WPG within errors.

\begin{figure}[t]
  \centering
  \includegraphics[]{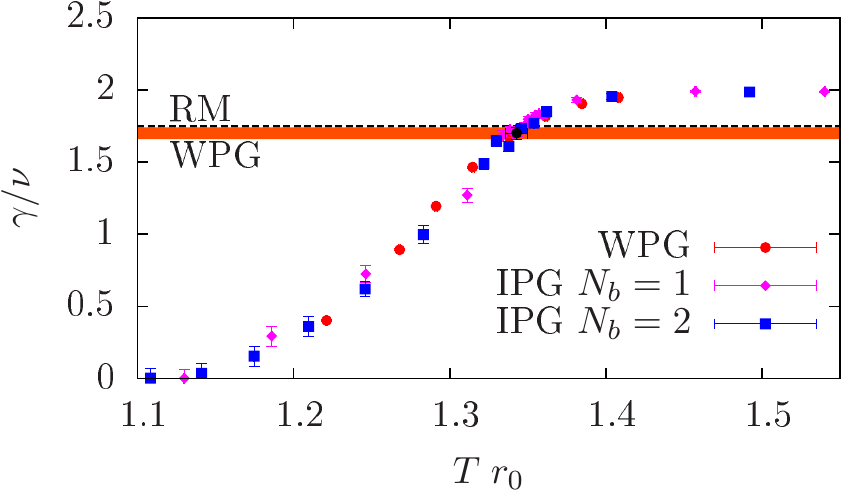} \hfill
  \includegraphics[]{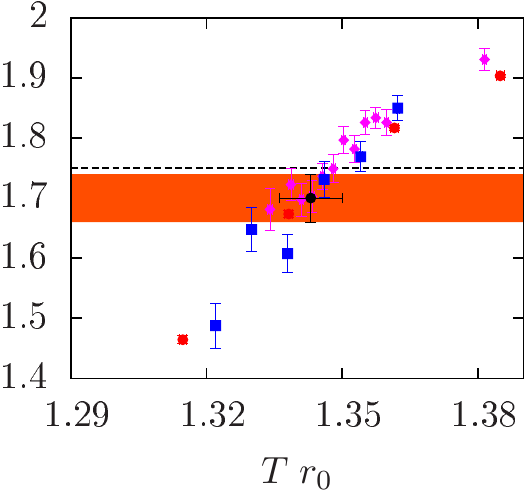}
  \caption{Results for $\gamma/\nu$ from the fits discussed in the text, for
    3d $\SU(2)$ gauge theory at $\Nt=4$.  The plot on the right displays
    the details of the $T_c(\infty)$ region.  The black dashed line is the
    prediction for $\gamma/\nu$ from a reduced model (RM)
    (cf.~\cite{Engels:1996dz}) and the black data point, together with the
    orange band, the result from the Bielefeld group~\cite{Engels:1996dz}.}
  \label{fig:fss-scaling-nt4}
\end{figure}%

\begin{figure}[t]
  \centering
  \includegraphics[]{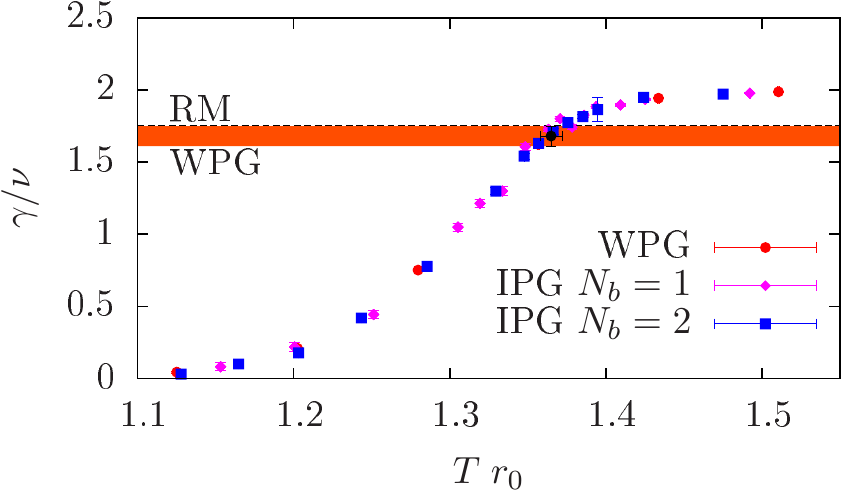} \hfill
  \includegraphics[]{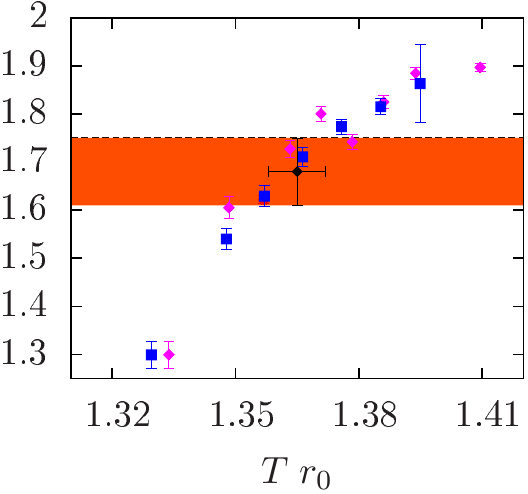}
  \caption{Same as figure~\ref{fig:fss-scaling-nt4} but for $\Nt=6$.}
  \label{fig:fss-scaling-nt6}
\end{figure}

The results of the analysis are shown in figures~\ref{fig:fss-scaling-nt4}
and~\ref{fig:fss-scaling-nt6} for $\Nt=4$ and 6, respectively. The plot on
the right in the figures shows the details in the region around
$T_c r_0(\infty)$. In the plots we do not show the results for $\gamma/\nu$ in
IPG since the uncertainties are rather large. However, when we assume that
a hypothetical high-precision result for $T_c(\infty)$ in IPG would be similar
to the result in WPG, the plot indicates that we would get a similar result for
$\gamma/\nu$, too. All in all we have compelling evidence
that the transitions in WPG and IPG theory are in the same universality
class. Moreover, cutoff effects are similar, so that we can expect this
statement to be true for the continuum limit as well.

\section{Conclusions}
\label{sec:conclusions}

In this paper we have tested the conjecture of the equivalence of
the continuum limit of induced pure gauge theory, formulated
originally by Budczies and Zirnbauer in~\cite{Budczies:2003za},
and pure gauge theory with Wilson's gauge action~\cite{Wilson:1974sk}.
To this end we have performed simulations with both discretizations
in three dimensions
with gauge group $\SU(2)$
at matched couplings to achieve similar lattice
spacings. The matching via the Sommer scale was discussed
in detail in section~\ref{sec:matching}. It is found to be in
good agreement with perturbation theory, both concerning its
functional form and the numerical results for the matching
coefficients (see also~\cite[sec. 4.3]{Brandt:2016duy} and
appendix~\ref{app:matching-pert}).

Using the matching, we have performed simulations at similar
lattice spacings for a high-precision comparison of
observables. In particular, we have looked at the leading and
subleading properties of the static $q\bar q$ potential at
intermediate and long distances in section~\ref{sec:qq-num}. The
leading properties are characterized by the string tension and the
subleading ones by the non-universal boundary coefficient \btt.
Both observables show excellent agreement between WPG and IPG
in the approach to the continuum for $\Nb=1$ and $\Nb=2$
and are similar already at finite lattice spacing. This indicates
that the continuum potential is identical, at least at the current
level of precision.

The thermodynamic properties of IPG theory have been investigated
in section~\ref{sec:finiteT} for two temporal extents, $\Nt=4$ and
6. Once more the agreement between the two theories for the
behavior of the Polyakov loop and its susceptibility is remarkable
for all volumes and over the full range of temperatures. Accordingly,
the ratio of critical exponents $\gamma/\nu$ agrees very well with
the known result for WPG from the Bielefeld group~\cite{Engels:1996dz},
indicating that the transition in both theories is in the same
universality class. Furthermore, the transition temperature in the
thermodynamic limit is also in agreement, which is another important
crosscheck since $T_c$ is a non-universal quantity.

One of the problems of the Kazakov-Migdal model~\cite{Kazakov:1992ym},
an earlier model supposed to induce Yang-Mills theory, is the existence
of a local center symmetry which constrains all Wilson lines to vanish.
The present model for induced Yang-Mills theory does not have this
problem. This is also reflected in the results for the Polyakov
loop obtained in section~\ref{sec:finiteT}. The results for the
Polyakov loop fall in the two sectors singled out by the center
symmetry of $\SU(2)$, i.e., they are located around $|L|$ and
$-|L|$ on the real axis. This shows that the model also
recovers the correct symmetry, whose breaking is associated with
the transition in Yang-Mills theory. As mentioned already in
section~\ref{sec:theory}, this is also expected from the symmetries
of the weight factor, eq.~\eqref{eq:wfact}.

Concerning numerical efficiency, our simulations of IPG theory
are up to a factor of 100 slower than the associated simulations
in WPG theory, depending on the bare parameters in the simulations.
We would like to stress that this inefficiency is
an attribute of the choice of the simulation algorithm alone, as
explained in detail in appendix~\ref{app:sims}. We are confident
that one can find an algorithm similar to the heat-bath algorithm
for WPG~\cite{Kennedy:1985nu}, which should then lead to a similar
performance. In particular, if
one is interested in simulations including fermionic degrees of
freedom, the Hybrid Monte Carlo (HMC)
algorithm~\cite{Duane:1987de} is supposedly the algorithm of
choice. We have tested the HMC algorithm for IPG theory in its
bosonized version and found it to perform almost as well as the
HMC for WPG theory in both $\SU(2)$ gauge theory for $d=3$ and
$\SU(3)$ in four dimensions. A general problem working in the bosonized version
is the increase in autocorrelations, which
can be enhanced up to an order of magnitude. However, this is
compensated to some extent by the speed-up of the individual HMC steps.
It is interesting to note that the HMC algorithm for IPG theory can
be set up in such a way that the auxiliary boson fields are drawn from
the exact distribution for the starting configuration. In this case
no communication is needed to evolve the link variables in pure gauge
theory, which results in a very efficient parallelization.

All in all, we find that the results from the two theories agree very
well already away from the continuum limit and that the agreement
improves as the continuum limit is approached. This is true already
for $\Nb=1$, which leads us to conclude that the bound in
eq.~\eqref{eq:equiv-bound} can indeed be relaxed for $d>2$. Our
results thus support the universality
argument given in \cite{Budczies:2003za}.
Therefore the modified BZ method, combined with a suitable
strong-coupling approach, can be used to reformulate lattice gauge
theories in a number of different ways.
We leave these reformulations for future work.

\section*{Acknowledgments}

The simulations have been done on the iDataCool and Athene clusters
at the University of Regensburg.
This work was supported by DFG in the framework of SFB/TRR-55.
B.B. has also received funding by the DFG via
the Emmy Noether Programme EN 1064/2-1.

\appendix

\section{Simulation details}
\label{app:sims}

\subsection{Update algorithm}
\label{app:update-algo}

The weight for a link $U_\mu(x)$ in the weight factor
\eqref{eq:wfact} in IPG theory is local so that we can use a
local version of the Metropolis algorithm~\cite{Metropolis:1953am}
for the simulations. More precisely, we propose a new link
$U'_\mu(x)$ and accept it with probability
\begin{align}
  \label{eq:accept_step}
  P = \min\Bigg\{ 1, \Bigg[ \prod_{p}
  \frac{\det\big(1-\frac{\alpha}{2}\big(U_{p}+U_{p}^\dagger\big)\big)}
  {\det\big(1-\frac{\alpha}{2}\big(U'_{p}+{U'_{p}}^\dagger\big)\big)}
  \Bigg]^{\Nb} \Bigg\} \,.
\end{align}
Here, the product over $p$ is taken over all plaquettes that include
the link $U_\mu(x)$, $U_{p}$ denotes the plaquette including the old
link $U_\mu(x)$, and $U'_{p}$ the plaquette including the new link
$U'_\mu(x)$. The crucial part affecting the efficiency of the
algorithm is to find new links $U'_\mu(x)$ that lead to a large
acceptance rate in the step defined by eq.~\eqref{eq:accept_step}.
Since in our pilot study of IPG theory efficiency of the algorithm was
not our primary concern, we simply took the new links $U'_\mu(x)$ to be
random $\SU(2)$ matrices in an $\epsilon$-surrounding of the old
links $U_\mu(x)$. In practice, we have generated a random element
$X=\sum_a x^a T^a\in\text{su}(2)$ (the Lie algebra of $\SU(2)$),
where the $T^a$ are the generators of $\SU(2)$ and the coefficients $x^a$
are taken from the interval $[-\epsilon,\epsilon]$, and constructed
the new link via $U'_\mu(x)=\exp(X)U_\mu(x)$. To make sure that in
each step a sufficient number of links is updated, we have tuned
$\epsilon$ so that the overall acceptance rate is around 80\%. To
further decorrelate two measurements we have separated them by
$N_\text{sw}$ sweeps, where $N_\text{sw}$ is chosen to be much larger
than the integrated autocorrelation time in units of lattice sweeps.

\subsection[Extraction of the static
\texorpdfstring{$q\bar{q}$}{qq} potential]{\boldmath Extraction of the
  static $q\bar{q}$ potential}
\label{app:r0-extract}

To compare the two theories we mostly use quantities that
are related to the potential between a static quark and antiquark.
The cleanest way, in terms of excited-state contaminations, to extract
the potential in numerical simulations is given by measuring the
correlation function of two Polyakov loops, $L(\vec{x})$, defined by
(cf. eq.~\eqref{eq:poly-loop})
\begin{align}
  \label{eq:single-poly-loop}
  L(\vec{x}) = \Tr \prod_{n_0=1}^{N_t}
  U_0(n_0\,a,\vec{x}) \,,
\end{align}
where $n_0$ denotes the temporal coordinate, $N_t$ is the number of
lattice points in the temporal direction, and $\vec{x}$ is the vector
including the spatial coordinates. The spectral representation of
a correlation function of two Polaykov loops separated by a distance
$R$ is given by
\be
\label{eq:Ploop-spect}
\ev{L^{\ast}(R)\:L(0)} = \sum_{n=0}^{\infty} b_n \: e^{-E_n(R)\:T}
\ee
for $R\ll L/2$, where $T=a N_t$ and $L=a N_s$ are the temporal and spatial
extents of the lattice, $b_i$ denotes the overlap between the operator and
the energy eigenstate, and $E_n(R)$ is the $n$-th energy level. Here the
energies are ordered in ascending order, i.e., $E_0<E_1<E_2<\ldots\;$. In
the limit $T\to\infty$ the right-hand side of eq.~\eqref{eq:Ploop-spect}
is dominated by the ground state with $E_0(R)=V(R)$. Excited states are
suppressed exponentially with $\exp\{-[E_i(R)-E_0(R)]T\}$. This means
that excited states can be neglected for large values of $T$. In this case
we can extract $V(R)$ via
\be
\label{eq:pot-extract}
V(R) = -\frac{1}{T} \ln\left[\ev{L^{\ast}(R)\:L(0)}\right] .
\ee

The correlation functions in eq.~\eqref{eq:Ploop-spect} suffer from a
well-known exponential decay of the signal-to-noise ratio with the area
enclosed by the two loops. This renders the extraction of the potential
difficult for large $R$. For simulations in WPG this problem can be
overcome by the use of a multilevel algorithm introduced by L\"uscher
and Weisz~\cite{Luscher:2001up}. The same algorithm can also be applied in
IPG since the locality properties of the action are similar to those of
the plaquette action. The details are discussed in
appendix~\ref{app:lw-algo}.

A suitable observable to set the scale with the static potential is
the Sommer parameter $r_0$~\cite{Sommer:1993ce}, which is defined
implicitly by
\be
\label{eq:sommer_para}
r_0^2F(r_0) = 1.65 \,,
\ee
where $F(R)$ is the force
\be
\label{eq:force}
F(R) \equiv \frac{\partial V(R)}{\partial R} \,.
\ee
To extract $r_0/a$ we use the following four methods (see
also~\cite{Brandt:2017yzw}):
\begin{enumerate}
 \item[(a)] a numerical polynomial interpolation of $R^2F(R)$,
 \item[(b)] a numerical rational interpolation of $R^2F(R)$,
 \item[(c)] a parameterization of the form~\cite{Sommer:1993ce}
 \be
  \label{eq:force-string}
  F(R) = f_0 + \frac{f_1}{R^2} + \frac{f_2}{R^4}
 \ee
 for the values of $R$ corresponding to the four nearest neighbors of $r_0$
 (motivated by the EST to LO),
 \item[(d)] the parameterization of \eqref{eq:force-string}
 with $f_2=0$ for the two nearest neighbors of $r_0$.
\end{enumerate}
The final estimate for $r_0/a$ is obtained from method (d), while methods
(a) to (c) are used to compute the systematic uncertainty associated with
the interpolation.

\subsection{Error reduction for large loops}
\label{app:lw-algo}

A suitable algorithm to overcome the exponential decay of the
signal-to-noise ratio for large loops is the multilevel algorithm
proposed by L\"uscher and Weisz~\cite{Luscher:2001up}.  The algorithm
relies on a key property of the theory, the locality of the
configuration weight, which ensures that sublattices, i.e., lattice
domains separated by a time slice with fixed spatial links, are
independent during local updates.  IPG theory also has this form of
locality since, as for the Wilson action, the weight for a particular
link $U_\mu(x)$ only depends on the plaquettes including
this link.  The error-reduction efficiency of the algorithm, however,
depends on the properties of the transfer matrix since this is the
object for which the uncertainty is decreased in the course of the
sublattice updates. If we are close enough to the continuum and both
theories indeed approach the same continuum limit, we can expect the
transfer matrix to be similar in both theories (given by the continuum
result plus lattice artifacts) so that the algorithm
should lead to a comparable error reduction also in the case of IPG
theory.

A way to estimate the amount of error reduction for Polyakov-loop
correlation functions, and thereby the optimal number of sublattice
updates for loops of a particular size, has been proposed
in~\cite{Majumdar:2002ga}.  Following this proposal we define the norm
$N(R)$ of a local two-link operator with link separation $R$ on a
particular sublattice as in~\cite[eq.~(4)]{Majumdar:2002ga}.  The
two-link operator plays the role of a transfer matrix for
Polyakov-loop correlators~\cite{Luscher:2001up}. The decay of $N(R)$
provides an estimate for the residual fluctuations of the transfer
matrix.  For a large number $n_t$ of sublattice updates we expect
$N(R)$ to fall off as $1/\sqrt{n_t}$.

\begin{figure}[t]
  \centering
  \includegraphics[scale=1.2]{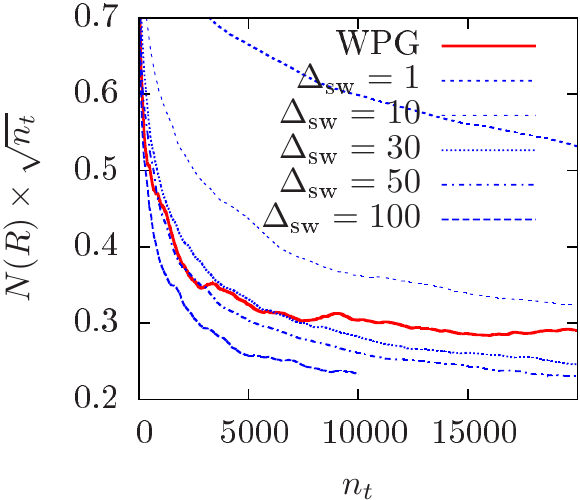}\hfill
  \includegraphics[scale=1.2]{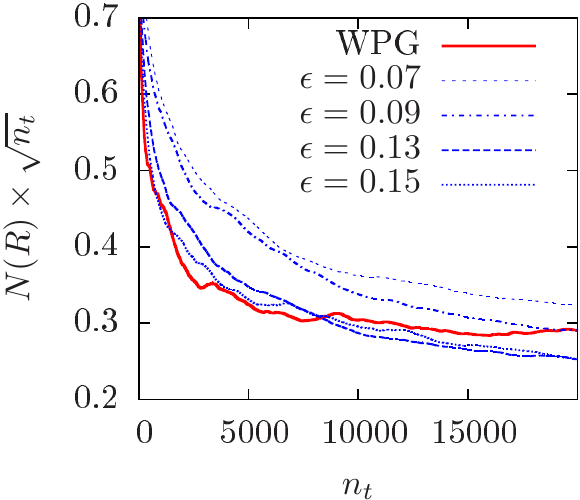}
  \caption{Results for the norm $N(R)$ from~\cite{Majumdar:2002ga} multiplied by
    the square root of the number of measurements in WPG and IPG with
    $\Nb=2$ for the cases $R=10$ and $\epsilon=0.07$
    (left) with different numbers $\Delta_\text{sw}$ of sublattice updates between two measurements,
    and $R=10$ and $\Delta_\text{sw}=15$ (right) with
    different values of $\epsilon$.}
  \label{fig:norm}
\end{figure}

To compare the error-reduction efficiency of the algorithm in the IPG and WPG
theories we have performed dedicated simulations in the $\SU(2)$
theory on a $24^3$ lattice at $\beta=5.0$ and the roughly equivalent
$\alpha=0.65$ in IPG with $\Nb=2$. We have fixed the
temporal extent $t_s$ of the sublattices to 2 throughout and measured two-link
operators with link separations between 2 and 10. In figure~\ref{fig:norm}
we show the results for link separation 10. The ``optimal'' value of
$n_t$ is taken to be the point at which the expected asymptotic behavior
$N(R)\sim 1/\sqrt{n_t}$ sets in for the largest value of $R$ considered.
After this point no further exponential error reduction is achieved.

The plot indicates that for WPG theory $n_t=15000$ to 20000 is
sufficient.  For IPG theory, $N(R)\times \sqrt{n_t}$ decreases
more slowly if we perform measurements at each sublattice update.
The main reason is that the sublattice configurations after one sublattice
sweep of local Metropolis updates in IPG theory are more correlated
than the configurations after one sublattice sweep of heat-bath updates for WPG
theory.  We therefore separate the measurements in IPG theory by
$\Delta_\text{sw}$ sublattice sweeps (which saves simulation time since the
measurement of the observable is more costly than a sweep of Metropolis
updates).  We see from figure~\ref{fig:norm} (left) that the decay of the
norm becomes stronger when we increase $\Delta_\text{sw}$. For the values shown
in figure~\ref{fig:norm} (left), where $\epsilon=0.07$,
$\Delta_\text{sw}=50-100$ appears to be a good choice. Moreover, the optimal
value of $\Delta_\text{sw}$ depends on the value of $\epsilon$, as shown in
figure~\ref{fig:norm} (right). Since in the final simulations we have used a
value of $\epsilon=0.14$, for which the integrated autocorrelation time is a factor
of 3 smaller than for $\epsilon=0.07$, the optimal value of $\Delta_\text{sw}$
can be taken to be around 20 for $\alpha=0.65$. A similar tuning can be done
for the other $\alpha$-values in the simulation. Since the optimal values of
$\epsilon$ and $\Delta_\text{sw}$ at a particular lattice spacing are a property of the
algorithm, we can use the same setup
for other values of $\Nb$ as long as we have tuned $\alpha$ so that we simulate
at similar lattice spacings. Note that the correctness of the algorithm does
not depend on the particular values of $\epsilon$ and $\Delta_\text{sw}$. The only effect of
a suboptimal tuning of $\epsilon$ or $\Delta_\text{sw}$ is less error reduction.

\section{Comparison of the matching with perturbation theory}
\label{app:matching-pert}

We want to compare the matching between the lattice couplings $\beta$ and
$\alpha$ discussed in section~\ref{sec:matching-r0} to the perturbative
results for the matching obtained in~\cite{Brandt:2016duy}. The perturbative
result (after analytic continuation to the region where $\alpha\to1$) as
given in~\cite[eq.~(4.15)]{Brandt:2016duy}, reformulated in terms of the
lattice couplings $\beta$ and $\alpha$, is given by
\be
\label{eq:pert-matching}
\beta = d_0(\Nb) \Nc \frac{\alpha}{1-\alpha} \Big( 1 + d_1(\Nb)
\frac{2(1-\alpha)}{d_0(\Nb)\,\alpha} + \Ord((1-\alpha)^2) \Big) .
\ee
The $\Nb$-dependence of $d_0(\Nb)$ and $d_1(\Nb)$ can be computed
in perturbation theory in the limit $\Nb\to\infty$. For a direct
comparison to perturbation theory it is convenient to replace the matching
function from eq.~\eqref{eq:mfit-2} by
\begin{align}
  \label{eq:mfit-pert}
  \beta(\alpha) = d_0(\Nb) \Nc \frac{\alpha}{1-\alpha} + \tilde{b}_0 +
  \tilde{b}_1\frac{(1-\alpha)}{\alpha} \,.
\end{align}
We are particularly interested in the coefficient $d_0$, for which the
perturbative prediction is given in~\cite[eq.~(4.78)]{Brandt:2016duy}.

To compare to perturbation theory we have used the parameterization
\eqref{eq:mfit-pert} for a comparison with the data for $\Nb=1$ and 2 from
section~\ref{sec:matching-r0} and performed additional simulations for
$\Nb=3$, 4 and 5, for which the simulation parameters are listed in
table~\ref{tab:simpoints_pmatch}. We observe that this parameterization works
well and leads to a similar description of the data as the parameterization
used in section~\ref{sec:matching-r0}. The results for $d_0(\Nb)/\Nb$, which have
already been shown in~\cite[Figure 2]{Brandt:2016duy}, are listed in
table~\ref{tab:matching-pcoef}.\footnote{We do not list the other
coefficients since we are mostly interested in a comparison to perturbation
theory, for which only $d_0$ is essential.} For the comparison itself we
refer to~\cite{Brandt:2016duy}.

\begin{table}[t]
  \centering
  \small
  \begin{tabular}{c|cc|ccccccc|l}
    \hline
    \hline
    $\Nb$ & $\alpha$ & size &
    $R$ & $t_s$ & $n_t$ & $\Delta_\text{sw}$ & meas & $N_\text{sw}$ &
    $\epsilon$ & \multicolumn{1}{c}{$r_0$} \\
    \hline
    \hline
    3 & 0.50 & $24^3$ & 1-9 & 2 & 100k & 20 & 2000 & 1000 &
    0.14 & 3.269(1)(1) \\
    & 0.60 & $36^3$ & 1-10 & 4 & 200k & 40 & 2000 & 1500 &
    0.10 & 5.349(1)(1) \\
    & 0.70 & $48^3$ & 1-11 & 6 & 500k & 100 & 1900 & 2000 &
    0.08 & 8.679(2)(1) \\
    \hline
    4 & 0.42 & $24^3$ & 1-10 & 2 &
    100k & 20 & 2000 & 1000 & 0.14 & 3.506(1)(1) \\
    & 0.51 & $36^3$ & 1-11 & 4 & 200k & 40 & 2000 & 1500 &
    0.10 & 5.417(1)(1) \\
    & 0.60 & $48^3$ & 1-11 & 6 & 500k & 100 & 1800 & 2000 &
    0.08 & 8.143(2)(1) \\
    \hline
    5 & 0.38 & $24^3$ & 1-9 & 2 &
    100k & 20 & 1900 & 1000 & 0.14 & 4.004(1)(4) \\
    & 0.49 & $36^3$ & 1-11 & 4 & 200k & 40 & 2000 & 1500 &
    0.10 & 6.733(2)(7) \\
    & 0.55 & $48^3$ & 1-12 & 6 & 500k & 100 & 2300 & 2000 &
    0.08 & 8.775(4)(1) \\
    \hline
    \hline
  \end{tabular}
  \caption{Simulation parameters and results for the measurements of
    $r_0$ in pure $\SU(2)$ IPG for $\Nb=3$, 4 and 5. Here, $R$ gives the
    range of $q\bar{q}$ separations used in the analysis of
    Polyakov-loop correlation functions, $t_s$ is the temporal extent
    of the L\"uscher-Weisz sublattices, $n_t$ is the number of sublattice
    updates, $\Delta_\text{sw}$ is the number of sweeps separating two
    sublattice measurements, $N_\text{sw}$ is the number of sweeps
    between two measurements, and $\epsilon$ is the size of the ball for the
    link proposal. For more details on the algorithms, e.g.,
    the choice of $\Delta_\text{sw}$ and $N_\text{sw}$, see 
    appendix~\ref{app:sims}.}
  \label{tab:simpoints_pmatch}
\end{table}

We can also compare directly to the matching results
of section~\ref{sec:matching-r0} by noting that the coefficients $b_{-1}$
and $d_0$ are related by
\be
\label{eq:d0-coeff}
d_0(\Nb) = b_{-1}(\Nb)/\Nc \,.
\ee
Using this relation to convert the results of eq.~\eqref{eq:beta_vs_alpha}
for $b_{-1}(\Nb)$ with $\Nb=1$ and 2 to results for $d_0$ gives
\be
\label{eq:d0-coeff-res}
d_0(1)/\Nb = 0.312(2) \quad \text{and} \quad d_0(2)/\Nb = 0.613(4) \,,
\ee
in perfect agreement with the results in table~\ref{tab:matching-pcoef}.

\begin{table}[h]
  \centering
  \small
  \begin{tabular}{l|ccccc}
    \hline
    \hline
    $\Nb$ & 1 & 2 & 3 & 4 & 5 \\
    \hline
    $d_0/\Nb$ & 0.311(2) & 0.614(3) & 0.735(3) & 0.793(3) & 0.847(6) \\
    \hline
    \hline
  \end{tabular}
  \caption{Numerical results for the perturbative coefficient
  $d_0(\Nb)$, normalized by $\Nb$.}
  \label{tab:matching-pcoef}
\end{table}


\providecommand{\href}[2]{#2}\begingroup\raggedright\endgroup

\end{document}